\documentclass[a4paper,11pt]{article}
\pdfoutput=1 

\usepackage{jheppub,braket} 

\usepackage[T1]{fontenc} 
\usepackage{longtable}
\DeclareMathOperator{\csch}{csch}
\usepackage{amsfonts,amsthm,amsmath,amssymb,graphicx,color}
\usepackage{xcolor,colortbl}
\usepackage{float}
\usepackage{graphicx}  
\usepackage{xcolor}
\usepackage{braket}
\usepackage{amssymb}   
\usepackage{amsmath,bm}
\usepackage{amsfonts}
\usepackage{bbm}
\usepackage{breqn}
\usepackage{setspace}
\usepackage{mathrsfs}

\definecolor{Gray}{gray}{0.95}
\definecolor{LightCyan}{rgb}{0.88,1,1}

\newcommand{\be}{\begin{equation}}
\newcommand{\ee}{\end{equation}}
\newcommand{\bea}{\begin{eqnarray}}
\newcommand{\eea}{\end{eqnarray}}

\newcommand{\Tr}{\text{Tr}}

\newcommand{\ba}{\begin{eqnarray}}
\newcommand{\ea}{\end{eqnarray}}

\newcommand{\prl}{Physical Review Letters}
\newcommand{\pra}{Physical Review A}
\newcommand{\prb}{Physical Review B}

\newcommand{\prd}{Physical Review D}

\newcolumntype{a}{>{\columncolor{Gray}}c}
\newcolumntype{b}{>{\columncolor{white}}c}

\title{\boldmath \texorpdfstring{$T\bar{T}$}{TEXT}, the entanglement wedge cross section, and the breakdown of the split property}

\author[a]{Meseret Asrat,}
\author[b]{Jonah Kudler-Flam}

\affiliation[a]{Enrico Fermi Institute, University of Chicago, IL~60637, USA}
\affiliation[b]{Kadanoff Center for Theoretical Physics, University of Chicago, IL~60637, USA}

\emailAdd{meseret@uchicago.edu}
\emailAdd{jkudlerflam@uchicago.edu}

\abstract{
We consider fine-grained probes of the entanglement structure of two dimensional conformal field theories deformed by the irrelevant double-trace operator $T\bar{T}$ and its closely related but nonetheless distinct single--trace counterpart.
For holographic conformal field theories, these deformations can be interpreted as modifications of bulk physics in the ultraviolet region of anti--de Sitter space. Consequently, we can use the Ryu-Takayanagi formula and its generalizations to mixed state entanglement measures to test highly nontrivial consistency conditions. In general, the agreement between bulk and boundary quantities requires the equivalence of partition functions on manifolds of arbitrary genus. For the single-trace deformation, which is dual to an asymptotically linear dilaton geometry, we find that the mutual information and reflected entropy diverge for disjoint intervals when the separation distance approaches a minimum, \textit{finite} value that depends solely on the deformation parameter.
This implies that the mutual information fails to serve as a geometric regulator which is related to the breakdown of the split property at the inverse Hagedorn temperature. In contrast, for the double-trace deformation, which is dual to anti-de Sitter space with a finite radial cutoff, we find all divergences to disappear including the standard quantum field theory ultraviolet divergence that is generically seen as disjoint intervals become adjacent. We furthermore compute reflected entropy in conformal perturbation theory. While we find formally similar behavior between bulk and boundary computations, we find quantitatively distinct results. We comment on the interpretation of these disagreements and the physics that must be altered to restore consistency. We also briefly discuss the $T{\bar J}$ and $J{\bar T}$ deformations.
}

\begin{document} 
\maketitle
\flushbottom


\section{Introduction}

The renormalization group is a fundamental concept in many-body physics and quantum field theory. It has been central to our understanding of the validity of particle physics, the emergence of macroscopic phenomena, and the space of quantum field theories. Starting from a fixed point of the renormalization group (a conformal field theory), one may perturb the action by a relevant operator, an interaction that becomes increasingly important as one flows to the IR. Relevant deformations compose the minority of terms that are available to add to the action, though their flows through theory space are generally well understood. On the contrary, the vast majority of possible deformations are irrelevant, describing the (generally infinite) UV directions that may flow to the same fixed point. Such deformations are usually difficult to understand with the infrared degrees of freedom, computational uncontrollable, and intractable.

A surprising and deeply consequential breakthrough came when Zamolodchikov and Smirnov showed that a particular class of irrelevant deformations of two dimensional  quantum field theories, in particular theories obtained by deforming via the determinant of the stress-energy tensor, are generally under control and ``solvable'' in the sense that the energy spectrum, partition functions, and (sometimes) the classical action may be computed exactly \cite{2004hep.th....1146Z, 2017NuPhB.915..363S,2016JHEP...10..112C,2018JHEP...06..149B}. Moreover, the UV descriptions of these theories are not local quantum field theories. The scale where locality breaks down is determined by the dimensionful deformation parameter, $\mu$. The understanding of this ``$T\bar T$'' deformation has seen tremendous progress including its reinterpretations as coupling the seed QFT to Jackiw-Teitelboim gravity \cite{2018JHEP...09..158D} or a flat, random metric \cite{2018JHEP...10..186C}. Furthermore, analogous irrelevant deformations have been put forward for theories with $U(1)$ currents that break Lorentz symmetry, the $T\bar J$ and $J \bar T$ deformations \cite{2018ScPP....5...48G}.

Due to the great success of the AdS$_3$/CFT$_2$ correspondence \cite{1999IJTP...38.1113M,2006hep.th....9074K}, a natural question to ask is how these deformations of two dimensional conformal field theories are viewed or interpreted 
from the bulk gravity (string) theory side. Thus far, gauge-gravity duality has mainly only been successful for theories that are asymptotically anti-de Sitter, corresponding to field theories that are controlled by a UV fixed point. Thus, one may hope, that these irrelevant deformations, which alter the UV physics, may guide us to understanding holography for quantum gravity theories with different asymptotics, in particular asymptotically flat space-times. There have been two\footnote{There is actually a third, recent proposal \cite{2020arXiv200306300H} motivated by Cardy's random geometry interpretation. We will return to this in the discussion section.} explicit proposals for how the asymptotics may be altered under the irrelevant deformations. 

The first proposal is what we refer to as ``cutoff AdS'' which has a surprisingly simple description \cite{2016arXiv161103470M}. Rather than the standard Dirichlet asymptotic boundary conditions of the GKPW dictionary \cite{1998AdTMP...2..253W,1998PhLB..428..105G}, one places Dirichlet boundary conditions at a finite radial cutoff. An important ingredient is that the deformation parameter is sign-definite in the direction where there always exist complex energies in the spectrum, a puzzling feature whose physical implications must be addressed. This conjecture has passed several tests such as matching the energy spectrum, thermodynamics, and signal propagation speeds \cite{2016arXiv161103470M}. However, in order to match bulk and boundary two-point functions, one must add an additional double-trace deformation to the boundary theory \cite{Kraus:2018xrn}. 

The other holographic proposal regards a closely related but distinct deformation of the conformal field theory, a ``single-trace'' $T \bar T$ \cite{2017arXiv170105576G,2017JHEP...12..155G,2018NuPhB.932..241A}. The following is meant by single-trace. We consider a symmetric orbifold theory $\mathcal{M}^N/S_N$\footnote{CFTs in the moduli space of this theory describe the long string sector of string theory on $AdS_3$ \cite{2001JMP....42.2929M, Argurio_2000}. ${\cal M}$ is the conformal field theory of a single long string.} (where $N$ is an integer). Rather than deforming the theory by the stress tensor of the full theory, we deform each block, $\mathcal{M}$, and take the symmetric product of the resulting theories. This proposal has the deformation parameter be sign-definite opposite to ``cutoff AdS.'' The energies of states (on a cylinder) are real. The bulk description of this deformation is a truly marginal deformation of the world-sheet string theory. This affects the bulk theory by changing the asymptotics to linear dilaton. Such asymptotics are quite close to asymptotically flat space-times though the high-energy density of states exhibit novel Hagedorn growth ($S\propto E$). Analogous to Ref.~\cite{2018ScPP....5...48G}, there are generalizations to the single-trace versions of the $T\bar J$ and $J \bar T$ deformations whose proposed holographic duals have warped AdS as asymptotics \cite{2019arXiv190500051C}. Also, see generalizations in Ref.~\cite{2019arXiv191112359A}.

Both of these proposals are fascinating and potentially quite important for our understanding of holography in generic space-times. It is important to both test these conjectures and to understand the novel structure of their dual (non-local) field theories. An illuminating observable is the entanglement entropy of subregions. Understanding entanglement structure has been central in characterizing many-body body systems such as gapped, critical, topologically ordered, and holographic systems \cite{2007JSMTE..08...24H,1994NuPhB.424..443H,2003PhRvL..90v7902V,2004JSMTE..06..002C,2009JPhA...42X4005C,2006PhRvL..96k0405L,2006PhRvL..96k0404K,2006PhRvL..96r1602R, 2006JHEP...08..045R, 2010GReGr..42.2323V,Hubeny:2007xt}. Presumably, the entanglement structure of these $T\bar T$ deformed theories will elucidate their properties. In particular, entanglement entropy has played a key role in our understanding of the renormalization group, providing c-functions in two and three dimensions that have clear information theoretic meaning regarding the number of degrees of freedom at each scale \cite{2004PhLB..600..142C,2012PhRvD..85l5016C,Liu:2012eea}. Generalized entropic c-functions in arbitrary dimensions for holographic theories were shown to hold in Ref.~\cite{2011JHEP...01..125M}.

Several studies have been conducted for entanglement entropy in holographic $T\bar T$ deformed theories \cite{2018PhRvL.121m1602D,2020JHEP...04..152L, 2018PhRvD..98h6025C, Banerjee:2019ewu,2019arXiv190603894J,2018NuPhB.935..290C,Murdia:2019fax,Gorbenko:2018oov,Park:2018snf,2019arXiv191104618A,2019arXiv190712603H,2019PhRvD..99j6008S,2019JHEP...11..171G}. While one can non-perturbatively compute the entropy using the Ryu-Takayanagi formula \cite{2006PhRvL..96r1602R, 2006JHEP...08..045R}, in general, one must use techniques from conformal perturbation theory to compute the entropy from the field theory side. One exception to this is for the entropy of very specific configurations such that the replica trick may be computed non-perturbatively. For example, for an entangling surface consisting of two antipodal points on a sphere, it is shown that the cutoff AdS and boundary computations precisely agree \cite{2018PhRvL.121m1602D}. Impressively, this technique has been generalized to the less symmetric case of a finite interval \cite{2020JHEP...04..152L}.

There are many reasons why the entanglement structure of $T\bar T$, $T{\bar J}$, and $J \bar T$ deformed theories deserves further attention, though the following three are our main motivations:
\begin{enumerate}
    \item Thus far, only von Neumann entropy has been studied which is only a reasonable measure of entanglement for pure states. This is severely limiting as there is deep structure in mixed state entanglement and multipartite entanglement, particularly in holographic systems. For this reason, we consider mixed state correlation measures such as the mutual information and measures dual to the entanglement wedge cross section, a bulk geometric object distinct from the Ryu-Takayanagi surface \cite{2018NatPh..14..573U,Nguyen:2017yqw, Kudler-Flam:2018qjo,Tamaoka:2018ned,2019arXiv190500577D}. We review the relevant holographic conjectures in the following subsection.
    \item The central input into the holographic dictionary is the equivalence of bulk and boundary partition functions
\begin{align}
     \mathcal{Z}_{\rm gravity}[\mathcal{B}] = \mathcal{Z}_{\rm QFT} [\partial \mathcal{B}],
     \label{adscftZ}
\end{align}
where $\mathcal{B}$ is an arbitrary bulk manifold and $\partial \mathcal{B}$ is its (asymptotic) boundary. This is the main ingredient in the derivation of the Ryu-Takayanagi formula \cite{2013JHEP...08..090L}. For the agreements in Refs.~\cite{2018PhRvL.121m1602D,2020JHEP...04..152L}, \eqref{adscftZ} only must hold for $\mathcal{B}$ a solid sphere. Such an equivalence of partition functions is quite unique. On the contrary, the replica tricks used for the mixed state correlation measures that we study implicitly require the equivalence of partition functions for $\partial \mathcal{B}$ arbitrary genus Riemann surfaces. This is a significantly stronger check of the holographic dualities. Surprisingly, we find that calculations on the two sides of the duality do not agree. 
\item While entanglement in local quantum field theory has been studied extensively, entanglement in non-local theories is much less understood. The $T\bar T$ deformation induces a flow that leads to non-locality at the UV scale. This provides us a rare tractable testing ground for studying information theoretic aspects of renormalization group flows and certain concepts in algebraic quantum field theory, such as nuclearity and the split property, that we review later in the introduction.
\end{enumerate}

The rest of the paper is organized as follows: In Section \ref{sec_lin_dilaton}, we introduce the asymptotically linear dilaton background proposed to be dual to the single-trace deformation. We generalize the results for von Neumann entropy in Refs.~\cite{2018NuPhB.935..290C,2019arXiv191104618A} to finite temperature states and compute the mutual information for disjoint intervals. Then, we compute the entanglement wedge cross section for disjoint intervals. We find that both quantities are UV divergent even when the intervals are only a finite distance away from each other determined by the non-locality scale of the deformed theory. Such divergences at finite distances signal breakdowns of the geometric regulators of Refs.~\cite{2015JHEP...10..003C,2019arXiv190500577D} that are used e.g.~to isolate $c$-functions, a consequence of the split property of quantum field theory \cite{Buchholz:1973bk, Doplicher:1984zz} ceasing to hold. Furthermore, both the mutual information and reflected entropy are monotonically increasing with the deformation parameter. 
In Section \ref{sec_cutoffAdS}, we do the same thing but for AdS with a hard finite radial cutoff at both zero and finite temperature. In sharp contrast to the previous section's results, we find the quantities to be UV finite even when the intervals are brought arbitrarily close together. Here, they are monotonically decreasing with (the absolute value of) the deformation parameter. 
In Section \ref{sec_CFT}, we perturbatively (in the deformation parameters) compute the reflected entropy for disjoint intervals. We find the first order corrections from the $J \bar T$ and $ T\bar J$ deformations to vanish due to twist fields being uncharged under the $U(1)$ symmetry. For double-trace $T \bar T$, we find formally similar, but numerically distinct, results to the holographic calculations, showing tension in the holographic proposal. In Section \ref{sec_discussion}, we discuss physical implications and open questions. Finally, in the appendices, we collect various derivations and formulas.

\subsection{Holographic entanglement and mixed-state correlation measures}

The Ryu-Takayanagi formula has become a standard in high-energy physics and its relation to quantum information theory \cite{2006PhRvL..96r1602R, 2006JHEP...08..045R}
\begin{align}
    S_A = \frac{1}{4G_N^{(d)}}\int_{\gamma_A} d^dx\sqrt{g}.
\end{align}
Here, $\gamma_A$ is the extremal surface (with respect to the integrand) homologous to boundary region $A$.
Important for our discussion is how this formula must be modified when the dilaton is not trivial.
In Ref.~\cite{2006JHEP...08..045R}, it was posited that in the case in which the dilaton is not constant
\begin{align}
    S_A = \frac{1}{4G_N^{(d)}}\int_{\gamma_A} d^dx e^{-2\Phi }\sqrt{g^{(s)}},
    \label{string_frame_RT}
\end{align}
where $\Phi$ is the dilaton and $g^{(s)}$ is the metric in string frame. We provide a simple derivation of this modification in Appendix \ref{app_string_frame}.

Various generalizations of the Ryu-Takayanagi formula have played important roles in recent years, such as a covariant description \cite{Hubeny:2007xt}, quantum corrections \cite{Faulkner:2013ana,Engelhardt:2014gca}, and R\'enyi entropies \cite{2016NatCo...712472D}, all of which have been derived under mild assumptions \cite{2013JHEP...08..090L, Faulkner:2013ana,2016NatCo...712472D,Dong:2016hjy,Dong:2017xht}.

\begin{figure}
    \centering
    \includegraphics[width = \textwidth]{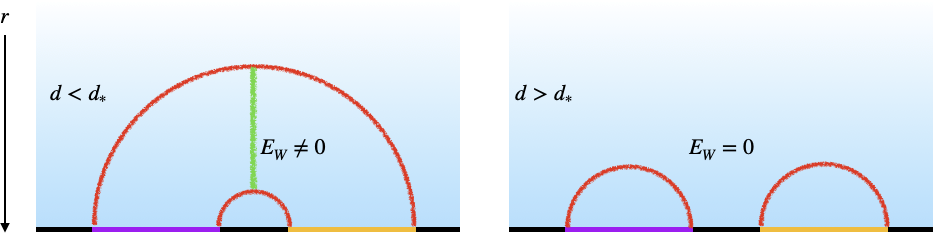}
    \caption{There are two phases of the entanglement wedge for disjoint intervals. In the connected regime (left), the Ryu-Takayanagi surfaces (red) stretch between the boundary two subregions (purple and orange). This phase manifestly has nontrivial mutual information and entanglement wedge cross section (green). Alternatively, when the intervals are sufficiently distant ($d > d_*$), the disconnected regime (right) dominates and the entanglement wedge becomes the union of the two individual entanglement wedges. This phase has manifest zero mutual information and entanglement wedge cross section.}
    \label{EW_transition_cartoon}
\end{figure}

A large generalization was proposed in the context of entanglement of purification (EoP), a mixed state correlation measure that reduces to the von Neumann entropy for pure states \cite{doi:10.1063/1.1498001}. Motivated by information-theoretic inequalities and tensor network models of holography, this was conjectured to be dual to the area of the entanglement wedge cross section \cite{2018NatPh..14..573U,Nguyen:2017yqw}. The entanglement wedge of boundary region $A$, $\Xi_A$, is the bulk codimension-one region whose boundary is $A \cup \gamma_A$. The entanglement wedge cross section of $A\cup B$, $E_W(A:B)$, is the extremal surface in $\Xi_{A\cup B}$ separating regions $A$ and $B$ (see Fig.~\ref{EW_transition_cartoon} for a depiction of $E_W$ for disjoint intervals). While well-motivated and certainly plausible, the $EoP=E_W$ conjecture is unlikely to be proven using known methods due to the large optimization procedure in the definition of EoP\footnote{Related optimized correlation measures were argued to be dual to $E_W$ in Refs.~\cite{Umemoto:2019jlz,PhysRevD.101.046015}}. The entanglement wedge cross section was later conjectured to be dual to logarithmic negativity \cite{Kudler-Flam:2018qjo}, a well-known mixed state entanglement measure that is only sensitive to purely quantum correlations \cite{1999JMOp...46..145E,2002PhRvA..65c2314V,2005PhRvL..95i0503P}. An important aspect of this proposal is that $E_W$ must backreact non-trivially on the geometry akin to the R\'enyi entropies \cite{2016NatCo...712472D}. This conjecture was later derived for AdS$_3$/CFT$_2$ in Ref.~\cite{Kusuki:2019zsp}. A similar quantity, the ``odd entropy,'' was conjectured to be dual to $E_W$ without backreaction in Ref.~\cite{Tamaoka:2018ned}. Finally, in Ref.~\cite{2019arXiv190500577D}, it was shown that $E_W$ was equal to half of the ``reflected entropy,'' which is the von Neumann entropy of a particular (not minimal) purification $\rho \rightarrow \ket{\rho^{1/2}}$. This was derived under mild assumptions in generic dimensions. 

Due to the fact that the reflected entropy conjecture does not require difficult backreaction in the bulk (like negativity) and it has a convincing derivation, in this paper, we focus on computing this quantity. However, due to its recent introduction in the literature, little is known about its properties i.e.~what is it really measuring\footnote{See discussion and analysis of this question in Refs.~\cite{2019arXiv190500577D,Kudler-Flam:2020url,Kusuki:2019rbk,Kusuki:2019evw,Akers:2019gcv,Longo:2019pjj}.}? Thus, we find it conceptually useful to consider it as a proxy for logarithmic negativity, whose quantum information theoretic properties are well understood. Precisely, in holographic theories, the negativity is equal to half of the R\'enyi reflected entropy at R\'enyi index $1/2$.

Here, we collect some properties of $E_W$ and the reflected entropy between two subsystems, denoted $A(A_i)$ and $B(B_i)$ below \cite{2018NatPh..14..573U,2019arXiv190500577D}.
\begin{enumerate}
    \item Reduction to von Neumann entropy: when computing $E_W$ for a bipartition of a global pure state, $E_W$ reduces to the standard Ryu-Takayanagi surface. Likewise, the reflected entropy reduces to twice the von Neumann entropy.
    \item Upper bound: $E_W(A:B)$ is always bounded from above by the entropies $S(A)$ and $S(B)$ and the inequality is only saturated for pure states. Similarly reflected entropy always is bounded by twice the entropies.
    \item Lower bound: $E_W(A:B)$ is always larger than half of the mutual information $I(A:B)$. Likewise, this holds for reflected entropy in all quantum systems $S_R(A:B) > I(A:B)$.
    \item Monotonicity: $E_W$ is monotonic under inclusions i.e.~$E_W(A, B) \leq E_W(A, B\cup C)$. This makes it a reasonable correlation measure. While this has been proven for the integer R\'enyi reflected entropies, it has yet to be proven in the von Neumann limit, though it is suspected to hold.
    \item Strong superadditivity: 
    $E_W$ obeys strong superadditivity $E_W(A_1\cup A_2, B_1 \cup B_2)\geq E_W(A_1, B_1)+ E_W(A_2, B_2)$. Interestingly, there are counterexamples to this property for the reflected entropy of classically correlated finite dimensional systems.
\end{enumerate}

\begin{figure}
    \centering
    \includegraphics[width = .6\textwidth]{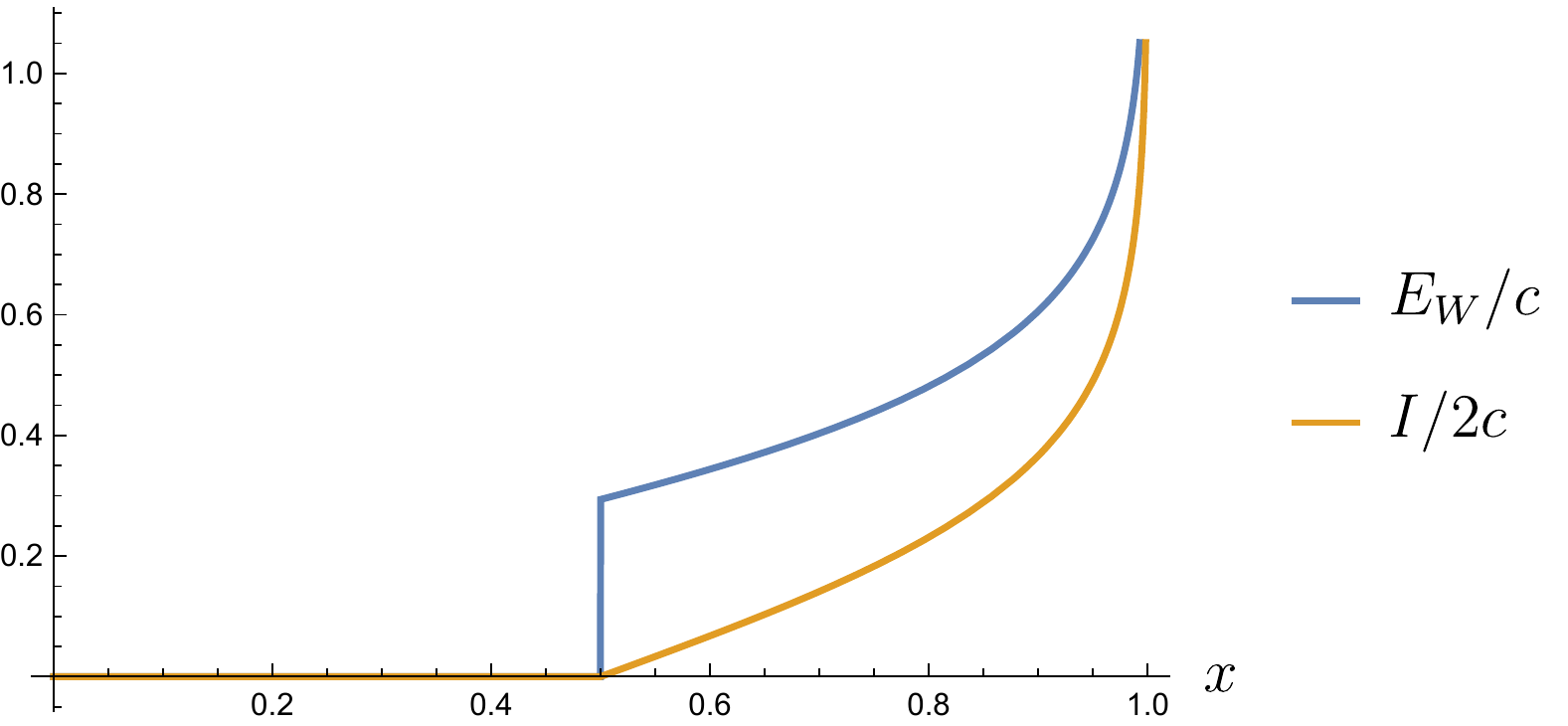}
    \caption{The entanglement wedge cross section and half the mutual information per central charge are plotted as a function of the conformally invariant cross-ratio \eqref{conf_cross} in the vacuum state of a 2D CFT. The bound $E_W > I/2$ is manifest. Notably, $E_W$ discontinuously jumps to zero at $x=1/2$. At this point, the mutual information is continuous but its first derivative is discontinuous. These analytic breakdowns are thought to be artifacts of the $c\rightarrow \infty$ limit.}
    \label{EW_intuition}
\end{figure}

To gain further intuition on $E_W$ and its relation/distinct to mutual information, we plot the two quantities for disjoint intervals ($[x_1, x_2]$ and $[x_3,x_4]$) in the vacuum of a conformal field theory in Fig.~\ref{EW_intuition}. Both quantities only depend on the conformally invariant cross-ratio
\begin{align}
    x = \frac{x_{21}x_{43}}{x_{31}x_{42}}.
    \label{conf_cross}
\end{align}
Much of this work is understanding how Fig.~\ref{EW_intuition} changes once we turn on the irrelevant deformations.

\subsection{Review of the split property of QFT and geometric regulators}

In this subsection, we provide a minimal review of the nuclearity condition and the split property of local quantum field theory. The interested reader may consult Ref.~\cite{Haag:1992hx} for further details. 

Unlike finite dimensional quantum systems, in quantum field theory, the Hilbert space does not admit a tensor factorization. The algebras of local observables in a subsystem of quantum field theory are generically type III von Neumann algebras. The partial trace of a state is thus ill--defined and consequently, the von Neumann entropy is always infinite. Even with these setbacks, one would like to be able to consider states localized in subregions and have well-defined correlation measures. 

While the von Neumann algebras, $\mathfrak{A}(K_r)$, of subregions, $K_r$, in quantum field theory are generically of type III, we can consider two disjoint causal diamonds, $K_1$ and $K_2$ with the causal complement of $K_2$, $\bar{K}_2$, subsuming $K_1$. The split property then asserts that there exists a type I factor, $\mathcal{N}$, such that
\begin{align}
    \mathfrak{A}(K_1) \subset \mathcal{N} \subset \mathfrak{A}(\bar{K}_2).
    \label{split}
\end{align}
Type I von Neumann algebras admit traces and are isomorphic to the set of bounded operators in some Hilbert space, so entropies may be well defined. To connect with our general intuition regarding tensor factorization of Hilbert spaces, we note that the split property equivalently says that there exists an isomorphism between $\mathfrak{A}(K_1)\vee \mathfrak{A}(K_2)$ and the tensor factorization $\mathfrak{A}(K_1) \otimes \mathfrak{A}(K_2)$. It is important to note that this mapping is not unique thence $\mathcal{N}$ is not unique. 

There is, however, a ``canonical'' choice for the split, $\mathcal{N}_{\psi}$, that always exists and was first discussed in Ref.~\cite{Doplicher:1984zz}. This is defined via the relation
\begin{align}
    \mathcal{N}_{\psi} = \mathfrak{A}(K_1) \vee J_{\psi} \mathfrak{A}(K_1) J_{\psi} ,
\end{align}
where $J_{\psi}$ is the modular conjugation operator of Tomita-Takesaki theory associated to the state $\psi$ and algebra $\mathfrak{A}(K_1) \vee \mathfrak{A}(K_2) $. It is the von Neumann entropy of the state in $\mathcal{N}_{\psi}$ that is associated to the reflected entropy. This is why the reflected entropy is so useful for us in understanding if/how the split property holds in a given theory for a given split distance, $s$. For more comprehensive discussion on this point, we direct the reader to Refs.~\cite{2019arXiv190500577D, Longo:2019pjj,2020arXiv200309546B}.

The split property has been shown to follow from the \textit{nuclearity condition} which we describe now. Consider the set
\begin{align}
    \mathcal{N}_{\beta,r} = e^{-\beta H}\mathfrak{A}^{(1)}(K_r)\ket{\Omega},
\end{align}
where $\ket{\Omega}$ is the vacuum state, $\beta > 0$, $H$ is the Hamiltonian, and $\mathfrak{A}^{(1)}(K_r)$ is the set of operators in $\mathfrak{A}(K_r)$ with norm less than or equal to $1$. This set is called nuclear if there exists a positive, trace class operator, $T_{\beta,r}$, such that
\begin{align}
    \mathcal{N}_{\beta,r} \subset T_{\beta,r} \mathcal{H}^{(1) },
\end{align}
where $\mathcal{H}^{(1) }$ are the elements of $\mathcal{H}$ with norm less than or equal to $1$. From here, we can define the \textit{nuclearity index} as
\begin{align}
    \nu_{\beta,r} = \inf_{ T_{\beta,r}} \Tr[ T_{\beta,r}],
\end{align}
which must be bounded as
\begin{align}
    \nu_{\beta,r} < e^{c r^d\beta^{-n}},
    \label{nuclearity_condition}
\end{align}
for some $c,n > 0$. $d$ is the number of spatial dimensions. It has been shown that if \eqref{nuclearity_condition} is satisfied, then the split property holds as long as $K_1$ and $K_2$ are disjoint, though they may be arbitrarily close \cite{buchholz1987}.

The nuclearity index is closely related to the thermodynamic partition function. In certain theories, the partition function can diverge at a finite temperature, $\beta_H$, called the Hagedorn temperature. This temperature governs the high-energy density of states. In these theories, the nuclearity condition is not satisfied and the split property fails to hold when $K_1$ and $K_2$ are at a finite distance from one another determined by $\beta_H$ (defined in \eqref{hagedorn_def}). $T\bar T$ deformed conformal field theories exhibit Hagedorn thermodynamics, so it is natural to investigate how this split property breaks down in these theories. This provides an opportunity where we have a concrete example to test the dynamical breakdown of \eqref{split}.

\begin{figure}
    \centering
    \includegraphics[height=5cm]{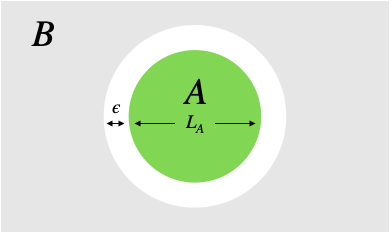}
    \caption{The geometric regularization scheme for the von Neumann entropy. We take $A$ and $B$ to be disjoint and separated by $\epsilon$. Then, the regulated ``entropy'' is $S_R(A:B)/2$ or $I(A:B)/2$ as $L_A/\epsilon \rightarrow 0$.}
    \label{regulatorfig}
\end{figure}

As previously mentioned, the von Neumann entropies associated to subregions in quantum field theory are generically divergent. It is thus desirable to regulate the entropy to extract quantities physically meaningful regarding the theory or specific state. Of particular interest are the quantities extracted from the von Neumann entropy that provide entropic $c$-functions \cite{2004PhLB..600..142C,2012PhRvD..85l5016C}. In particular, the von Neumann entropy for a ball-shaped region, $A$, has a generic expansion including the following terms
\begin{align}
    S_A \supset \begin{cases}
    (-1)^{\frac{d}{2}-1}4 c_0 \log\left[ \frac{L_A}{\epsilon_{UV}}\right], & d \in 2\mathbb{Z}
    \\
    (-1)^{\frac{d}{2}-1}2\pi c_0 , & d \in 2\mathbb{Z}+1
    \end{cases},
\end{align}
where $L_A$ is the radius of region $A$ and $\epsilon_{UV}$ is the ultraviolet cutoff. The constants $c_0$ are universal and independent of the details of the UV regularization. The definition of $c_0$ can be made unambiguous if the UV regulator is chosen appropriately as to allow the divergent terms in the entropy to be expressible as ``geometric,'' i.e.~integrals over the entangling surface \cite{2011PhRvB..84s5120G,Liu:2012eea}. The validity of the split property hints that the mutual information and reflected entropy of nearby, but disjoint regions are natural ``geometric'' regulators \cite{2015JHEP...10..003C,2019arXiv190500577D}. Both of these quantities may be well-defined without assuming a tensor factorization. The regularization scheme is imposed by considering region $A$ and its causal complement with the region at their interface of width $\epsilon$ excised (see Fig.~\ref{regulatorfig}). Then, one should take the limit of $L_A/\epsilon \rightarrow 0$ , though it is important that during this limit, one keeps $\epsilon \gg \epsilon_{UV}$. Then, in the final expression, all constants, e.g.~$c_0$, are physical and unambiguous. We will see that the ability of taking the $L_A/\epsilon \rightarrow 0$ limit disappears when we consider $T\bar T$ deformed theories.

\section{Single-trace \texorpdfstring{$T\bar T$}{TEXT} and the linear dilaton background}
\label{sec_lin_dilaton}

Let us begin by reviewing the holographic proposal of Refs.~\cite{2017arXiv170105576G,2017JHEP...12..155G,2018NuPhB.932..241A,2019arXiv190500051C}. This involves deforming the worldsheet conformal field theory by a linear combination of truly marginal current-current operators
\begin{align}
    \delta \mathcal{L}_{ws} = \lambda J^{-}_{SL}\bar{J}^{-}_{SL} + \lambda_+K\bar{J}^{-}_{SL}+ \lambda_-\bar{K}J^{-}_{SL},
\end{align}
where $K$ ($\bar{K}$) are worldsheet $U(1)$ currents associated to left (right)-moving momenta on a unit circle, and $J^{-}_{SL}({\bar{J}^{-}_{SL}})$ are bosonic $SL(2, R)$ left (right)-moving worldsheet affine currents.

It was shown that this deformation in the long string sector is equivalent to deforming ${\cal M}$ by a general linear combination of the irrelevant operators
\begin{align}
    \delta \mathcal{L}  = -\mu T \bar{T} - \mu_+ J \bar{T} - \mu_- \bar{J} T.
\end{align}
where $J$ and ${\bar J}$ are left and right moving U(1) currents, respectively, and $T$ and ${\bar T}$ are the holomorphic and anti–holomorphic stress tensor components, respectively, of the conformal field theory $\cal M$.
With our normalization, the coupling constants of the spacetime and worldsheet deformations are related as\footnote{Our $\mu$ is equal to $-{\mu\over 4\pi^2}$ from Ref.~\cite{2016arXiv161103470M}.}
\begin{align}
    \mu= \frac{6\alpha'\lambda}{c\pi} , \quad \mu_{\pm} = {4\over \pi}\sqrt{\frac{3\alpha'}{ c}}\lambda_{\pm},
    \label{parameter_relations}
\end{align}
where $\alpha'$ is the Regge slope, and $\mu,\ \mu_\pm > 0.$

Suppressing internal compact dimensions, the bulk background induced by this deformation has string frame metric \cite{2019arXiv191104618A,Araujo:2018rho,2019arXiv190500051C}
\begin{align}
    \label{lin_dil_metric}
    ds^2 = d\phi^2 + hd\gamma d\bar{\gamma} + \frac{2h\lambda_+}{\sqrt{k}}d\psi d\bar{\gamma}+ \frac{2h\lambda_-}{\sqrt{k}}d\psi d\bar{\gamma}+ \frac{1}{k}hf^{-1}d\psi^2,
\end{align}
where the dilaton $\Phi$, and Neveu--Schwartz two–form $B$ are
\begin{align}
    e^{2\Phi} &= g_s^2 e^{-2\phi}h, \quad B_{\gamma \bar{\gamma}} = g_{\gamma \bar{\gamma}}, \quad  B_{\gamma \psi} = g_{\gamma \psi}\quad  B_{\psi \bar{\gamma}} = g_{ \bar{\gamma} \psi},
\end{align}
and 
\begin{align}
    h^{-1} &= e^{-2\phi}+ \lambda -4\lambda_+ \lambda_-, \quad f^{-1} = h^{-1} + 4\lambda_+\lambda_-.
\end{align}
$\psi$ is the coordinate on the circle $S^1$, $\psi \sim \psi + 2\pi$. The boundary is located at $\phi = +\infty$. The coordinates $\gamma$ and $\bar\gamma$ are
\begin{equation}
    l_s\gamma = t + x, \quad l_s\bar\gamma = -t + x, 
\end{equation}
where $l_s = \sqrt{\alpha'}$ is the intrinsic string length.

We restrict to the parameter range where the energy spectrum was shown to be real and that there are no closed time like curves \cite{2019arXiv190500051C}
\begin{align}
    \frac{\lambda}{4\pi \alpha'}- \frac{(\lambda_+ + \lambda_-)^2}{32\alpha'} > 0.
\end{align}
Furthermore, we take $\lambda_+ = \lambda_- = {\frac{\delta}{ 2}}$.
\begin{figure}
    \centering
    \includegraphics[width = .6\textwidth]{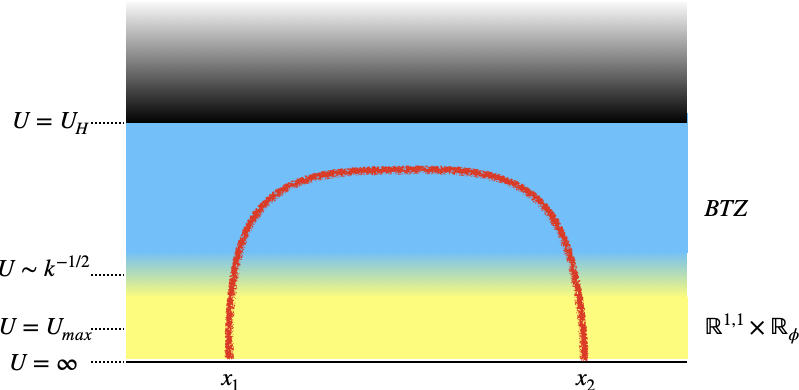}
    \caption{The single-trace $T \bar{T}$ deformation changes the bulk AdS$_3$ geometry to asymptote in the UV to a linear dilaton regime represented in yellow. This smoothly crosses over to the undeformed AdS$_3$ regime in the IR. For further generality, we have placed a black hole in the IR regime which sets the boundary theory to finite temperature. The IR region is thus more generally that of the BTZ black hole \cite{1992PhRvL..69.1849B}.}
    \label{linear_dilaton_BH_geometry_cartoon}
\end{figure}

We show the corresponding geometry in Fig.~\ref{linear_dilaton_BH_geometry_cartoon} for the case in which $\delta = 0$. The linear dilaton regime controls the UV, while vacuum AdS$_3$ controls the IR. They are smoothly connected at a scale determined by $k^{1/2}$ where $k$ is the level of the worldsheet $SL(2,\mathbb{R})$ algebra related to the AdS radius as $l_{AdS} = \sqrt{k\alpha'}$. Crucially, there is a radial (renormalization) scale where local physics breaks down, $U_{max}$. This is determined by the deformation parameter. The length scale in the spacetime CFT where this breaks down is\footnote{For the special case that $\mu = \mu_+ + \mu_-$, $l_{min}$ is doubled.}
\begin{align}
    l_{\rm min} = \frac{ \pi}{2}\sqrt{ \frac{c\pi \mu}{6}} = \frac{\beta_H}{4}. 
\end{align}
$\beta_H$ is the Hagedorn temperature governing the asymptotic (${E\rightarrow \infty}$) density of states
\begin{align}
    S = \beta_H E.
    \label{hagedorn_def}
\end{align}
Any observables in the field theory probing shorter distances than $l_{min}$ cease to make sense. We will see that this non-locality scale plays an important role for the mutual information and reflected entropy. We stress that all results in this section and Section \ref{sec_cutoffAdS} are computed from the gravity side and are thus contingent on the validity of the holographic dualities, the RT formula, and its generalizations.



\subsection{Vacuum}\label{sec_vacuum}

\subsubsection{Mutual Information}
In this subsection, we study the mutual information for two disjoint intervals of lengths $l_A$ and $l_B$ separated by a distance $d$. We begin with the case in which $\delta = 0,\ \lambda = 0$ corresponding to the vacuum of a conformal field theory. In this case, the mutual information takes the well-known form \cite{2010PhRvD..82l6010H}
\begin{align}
    I = \max\left[\frac{c}{3}\log\left(\frac{l_A l_B}{d(l_A + l_B + d)} \right),0\right].
\end{align}
where $c$ is the Brown–Henneaux central charge \cite{Brown:1986nw}. The mutual information is independent of the ultraviolet cutoff. 
The critical distance where the phase transition between connected and disconnected entanglement wedges occurs is given by 
\begin{align}
    d_* = \frac{1}{2}\left( \sqrt{l_A^2 + l_B^2 + 6l_A l_B} - l_A-l_B\right).
\end{align}

Under the assumption that the minimal surfaces do not break the symmetries of the spacetime metric, the von Neumann entropy of single intervals were evaluated using the Ryu-Takayanagi formula in Ref.~\cite{2019arXiv191104618A}. In this section, we make the trivial generalization of these results to the holographic mutual information of disjoint intervals, though we find intriguing new physics. 

In the rest of this subsection we consider the case in which $\delta = 0,\ \lambda \neq 0$. In this case, the entropy and the interval length are given by \cite{2019arXiv191104618A}\footnote{We can invert the equation for the interval length to write $\alpha$ as a function of $l, c, \mu$, and we can use this in the entropy to get an expression that only depends on $l, c, \mu$ and the UV cutoff $\epsilon_{UV}$. Once we get the entropy, we can use it to compute the mutual information.}
\begin{equation}
    S(\alpha) = {c\over 3}{\frac{1}{\alpha + 1}}\left[\left(2\alpha - \alpha^2{\frac{d}{ d\xi}}\right)\left.{1\over \xi + 1}\Pi(\varphi, n, k)\right|_{\xi = 0} + F(\varphi, k)\right],
\end{equation}
\begin{equation}
    {l \over l_{\rm min}} = {\pi\over 4}\sqrt{1+\alpha\over \alpha}E(\varphi, k),
\end{equation}
where
\begin{align}
\varphi = \arcsin{\sqrt{\left({ \alpha + 1\over 2\alpha + 1}\right) \left({1 + {3\epsilon_{UV}^2\alpha\over 2c\pi\mu}}\right)}},\quad n = {2\alpha + 1\over   \alpha + 1}\cdot { 1 \over \xi  + 1}, \quad k = \sqrt{{ 1\over  \alpha + 1}\cdot {2 \alpha + 1\over \alpha + 1}}   .
\end{align}
$F(\varphi, k)$, $\Pi(\varphi, n, k)$, and $E(\varphi, k)$ are the incomplete elliptic integrals of the first, second, and third kinds respectively. In our conventions, they are defined as
\begin{equation}
    \Pi(\varphi, n, k) = \int_0^\varphi {d\theta\over (1 - n\sin^2\theta)\sqrt{1 - k^2\sin^2\theta}}, \quad E(\varphi, k) = \int_0^\varphi d\theta \sqrt{1 - k^2\sin^2\theta},
\end{equation}
and $F(\varphi, k) = \Pi(\varphi, 0,k)$.

\begin{figure}
    \centering
    \includegraphics[width = .48\textwidth]{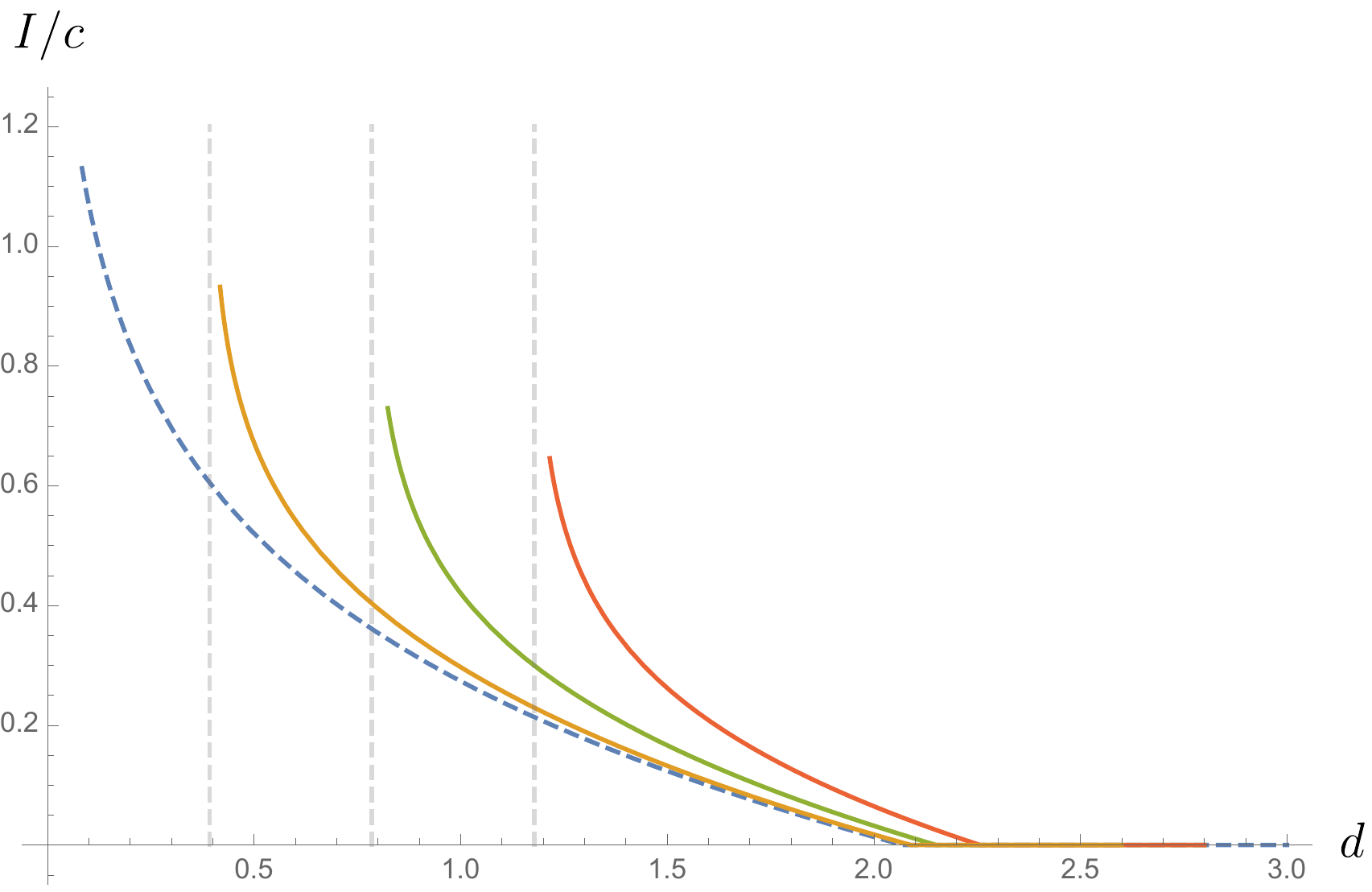}
    \includegraphics[width = .48\textwidth]{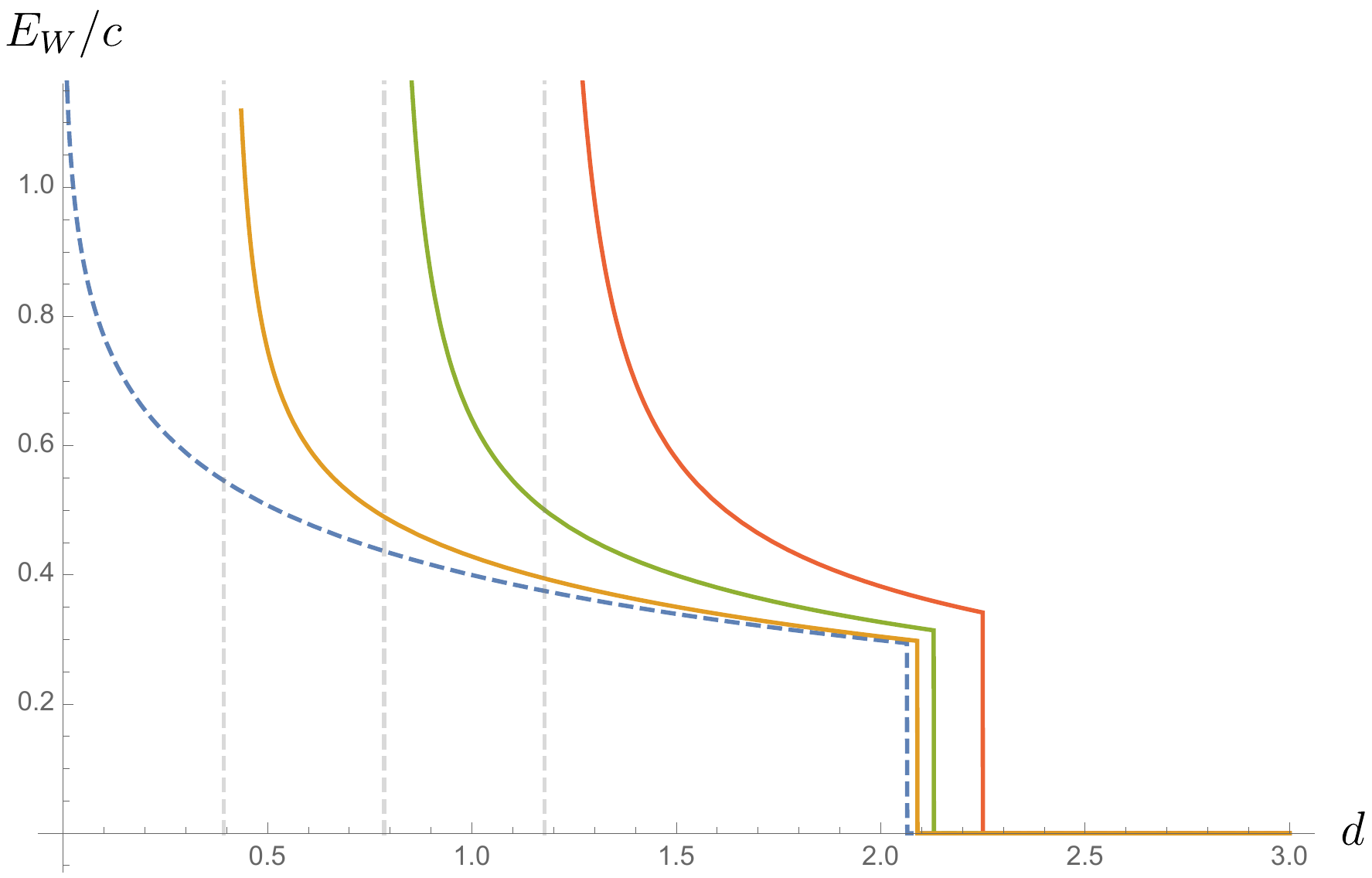}
    \caption{Left: the mutual information per central charge for disjoint intervals in the vacuum. Each subregion is of length $5$ and we plot $\sqrt{\frac{\pi c \mu}{6}} = \{0, \frac{1}{4}, \frac{1}{2}, \frac{3}{4} \}$ (blue to red respectively). We mark, with vertical dashed lines, the corresponding values of $l_{min}$ where the mutual information intriguingly diverges. Right: the area of the entanglement wedge cross section for the same parameters. We find the same divergences at $d= l_{min}$.}
    \label{linear_dilaton_MI_varyd}
\end{figure}
We have introduced the UV cutoff, $\epsilon_{UV}$, since the entropy $S$ is UV divergent. The mutual information
\begin{align}
     I(A,B) = S(l_A, \lambda) + S(l_B, \lambda) -\min [S_A(l_A, \lambda)+S(l_B, \lambda), S(l_A+l_B+d, \lambda) + S(d, \lambda)] .
     \label{mutual_info}
 \end{align}
is independent of the UV cutoff $\epsilon_{UV}$. However, we find that a divergence emerges for the mutual information even when the intervals are separated by a finite distance. At short distances, for intervals of equal length $l$, it diverges as $I\propto (d-l_{\min})^{-1}$ (Fig.~\ref{linear_dilaton_MI_varyd}). The same divergence was also noted for the entropic c--fucntion in \cite{2019arXiv191104618A,2019arXiv190500051C} at short distances.  This provides a fascinating breakdown of the split property of quantum field theory and the mutual information regulator of Ref.~\cite{2015JHEP...10..003C}. 
 



\subsubsection{Reflected entropy}

In this subsection, we consider the entanglement wedge cross section for the intervals. This has been computed, for example, in the vacuum state of an undeformed conformal field theory \cite{2018NatPh..14..573U}
\begin{align}
    E_W = \frac{c}{6} \log \left(\frac{1 + \sqrt{x}}{1-\sqrt{x}} \right), \quad x = \frac{l_Al_B}{ (l_A+d) (l_B+d)}.
    \label{EW_dis_CFT}
\end{align}
This is a UV finite quantity and, as previously discussed, has been proposed as a natural regulator for the von Neumann entropy which is universally UV divergent in local quantum field theory \cite{2019arXiv190500577D}\footnote{See related work in Refs.~\cite{1994RvMaP...6.1127N,2007arXiv0712.4403S,2010SHPMP..41..104S,NARNHOFER2002111,Narnhofer_2011,2018AnHP...19.1817O,2020arXiv200309546B,2020CMaPh.tmp...62L}.}. One advantage of using the reflected entropy as a regulator instead of the mutual information is that it is actually an \textit{entropy}. However, like the mutual information regulator, we will see that this also breaks down when we turn on the deformations (Fig.~\ref{linear_dilaton_MI_varyd}).

For simplicity, we will consider the case where $l_A = l_B \equiv l$ so that the minimal entanglement wedge cross section is purely radial in the bulk metric. 
Taking into account the factor of the dilaton in \eqref{string_frame_RT},
\begin{align}
   E_W =\frac{\sqrt{k\alpha'}}{4G_N^{(3)}}\int_{\phi_{-}}^{\phi_{+}}\frac{e^{2\phi}}{h} d\phi = \frac{c}{12}\left[\log\left({\alpha_{\rm +}\over \alpha_{\rm -}}\right) + \chi\left(\alpha_{\rm +} - \alpha_{\rm -}\right)\right], \quad \chi = 1 - {\delta^2\over \lambda},
   \label{EW_deformed}
\end{align}
where $\alpha_{-}$ and $\alpha_{+}$ are related to the turning points of the two minimal surfaces corresponding to the intervals of lengths (along the non-compact direction, $x$) $2l + d$ and $d$ respectively. They are related by \cite{2019arXiv191104618A}
\begin{align}
    \frac{2l + d}{l_{\rm min}} &={\pi\over 4}\sqrt{ \frac{1+\alpha_{\rm -}}{\alpha_{\rm -}}}E\left(\arcsin \sqrt{\frac{1+\chi\alpha_{\rm -}}{1+2\chi\alpha_{\rm -}}}, \sqrt{\frac{1+2\chi \alpha_{\rm -}}{\left( 1+\chi \alpha_{\rm -}\right)\left(1+\alpha_{\rm -}\right)}} \right),
\\
    \frac{d}{l_{\rm min}} &={\pi\over 4}\sqrt{ \frac{1+\alpha_{+}}{\alpha_{\rm +}}}E\left(\arcsin \sqrt{\frac{1+\chi\alpha_{\rm +}}{1+2\chi\alpha_{\rm +}}}, \sqrt{\frac{1+2\chi \alpha_{\rm +}}{\left( 1+\chi \alpha_{\rm +}\right)\left(1+\alpha_{\rm +}\right)}} \right).
\end{align}
These equations can be inverted to obtain equations for $\alpha_{\rm -}$ and $\alpha_{\rm +}$ in terms of $l, d, c, \mu, \mu_\pm$. Once we have these equations, we can use them to compute the wedge cross section.
Note that when $\delta = 0, \ \lambda = 0$, this reproduces \eqref{EW_dis_CFT}.





The coupling $\delta$ appears as $\delta^2$ and therefore, in perturbation theory, the ${\cal O}(\delta)$ term is zero. We will briefly discuss this case in perturbation theory later tn the paper from the boundary field theory side. In contrast, the ${\cal O}(\lambda)$ contribution is nonzero. This comes from the second, non-logarithmic term in \eqref{EW_deformed} giving
\begin{align}
    E_{W} = \frac{c}{6}\log\left(\frac{2l + d}{d}\right)  +{4\alpha'\lambda c l (l + d)\over 3 d^2 (d + 2l)^2} +{\cal O}(\lambda^2).
\end{align}
Note that the non-logarithmic term exists because the dilaton is not a constant. Using \eqref{parameter_relations}, we find 
\begin{align}
    E_{W} = \frac{c}{6}\log\left(\frac{2l + d}{d}\right)  +{2\pi c^2\mu l (l + d)\over 9 d^2 (d + 2l)^2} +{\cal O}(\mu^2).
    \label{zero_temp_lin_dil_EW_pert}
\end{align}
Because $\mu > 0$, this signals an \textit{increase} in the correlations between the subregions. The non--perturbative result \eqref{EW_deformed} is depicted in (Fig.~\ref{linear_dilaton_MI_varyd}) with $\delta = 0$. It is non--decreasing along the RG towards the UV and it diverges at short distances as the separtion distance $d$ approaches the non--locality scale $l_{\rm min}$.

\subsection{Finite temperature}

It is interesting to consider the generalization of the results obatined at zero temperature to finite temperature. Thermal states are holographically dual to black holes. Roughly, increasing the temperature decreases the ``quantumness'' of the state and quantum correlations are destroyed at length scales larger than the inverse temperature $\beta$. We are able to verify this explicitly by writing the generalization of the asymptotically linear dilaton metric (with $\delta = 0$) to include black hole solution. The spatial metric is
\begin{align}
    ds^2 &= \alpha' f d\phi^2 + hdx^2,
    \label{dilaton_BH_metric}
    \\
    f^{-1} &= 1 - e^{2(\phi_{ H} - \phi)}, \quad h^{-1}(\phi) = \lambda + e^{-2\phi},
\end{align}
where $\phi_H$ is determined by the temperature as
\begin{align}
     \beta = {2\pi\sqrt{\alpha'h^{-1}(\phi_{ H})}}.
\end{align}
This metric describes a BTZ black hole deep in the IR and linear dilaton asymptotics in the UV as was shown in Fig.~\ref{linear_dilaton_BH_geometry_cartoon}.

In what follows, we generalize the computations of Ref.~\cite{2019arXiv191104618A} and subsection \ref{sec_vacuum} of this paper for finite temperature with $\delta = 0$. We leave the case in which $\delta \neq 0$ for future work. In this paper, we work on the black hole background \eqref{dilaton_BH_metric}.

\subsubsection{Mutual information}

In this subsection, we study the mutual information for intervals of lengths $l_A$ and $l_B$ separated by a distance $d$.



We first compute the von Neumann entropy for an interval of length $l$. In Ref.~\cite{2018NuPhB.935..290C}, two distinct surfaces were considered as saddle points, the ``connected surface,'' which is the standard surface shown in Fig.~\ref{linear_dilaton_BH_geometry_cartoon}, and the ``disconnected surface'' which consists of two disconnected radial geodesics that terminate on the horizon of the black hole. The Ryu-Takayanagi prescription tells us to take the minimum of these two saddles. While this disconnected surface is natural to consider for confining geometries that have compact dimensions ``capping off'' at the horizon because a zero area tube may connect them (e.g.~Refs.~\cite{2007JHEP...01..090N, 2008NuPhB.796..274K}), this surface does not obey the homology constraint for the theories we study, so it should not be considered. In particular, the $\lambda, \delta \rightarrow 0$ limit should reproduce the universal CFT result
\begin{align}
    S = \frac{c}{3}\log \left(\frac{\beta}{\pi \epsilon_{UV}}\sinh\left(\frac{\pi l}{\beta} \right) \right),
    \label{univ_thermal_ent}
\end{align}
and including the disconnected surface in this limit causes the bulk and boundary computations to disagree. For these reasons, we only consider the connected regime for the von Neumann entropy.

The connected entanglement entropy is given by the integral
\begin{align}
    S_C = {\sqrt{k\alpha'}\over 4G_{\rm N}^{(3)}}\int_1^{x_\infty} dx\sqrt{\alpha x + 1\over (x - x_{H})(x - 1)(\alpha x + \alpha + 1)}(\alpha x + 1),
\end{align}
which is UV divergent. We solve for the entanglement entropy in closed form 
\begin{align}
 S_C = 
{c\over 3}\left\{{1\over \sqrt{(\alpha + 1)(\gamma + \alpha + 1)}}\left[\left(2\alpha - \alpha^2{d\over d\xi}\right)\right.\right.
\left.\left({\gamma\over \xi \gamma + \alpha}F(\varphi, k) + {\gamma + \alpha\over (\xi \gamma + \alpha)(\xi + 1)}\Pi (\varphi, n, k)\right)\right|_{\xi = 0}
\nonumber
\\
+ \left.\left.F(\varphi, k)\right]\right\},
\label{von_entropy}
\end{align}
where \cite{PFB:1971} 
\begin{align}
\varphi = \arcsin{\sqrt{{\gamma + \alpha + 1\over 2\alpha + 1}\cdot {1 +{3L_\Lambda^2 \alpha\over 2c\pi\mu}\over 1 +{3L_\Lambda^2 \gamma\over 2c\pi\mu}}}},\quad n = {2\alpha + 1\over \gamma + \alpha + 1}\cdot {\xi \gamma + \alpha \over \xi \alpha + \alpha}, \quad k = \sqrt{{\gamma + 1\over \gamma + \alpha + 1}\cdot {2 \alpha + 1\over \alpha + 1}},    
\end{align}
and
\begin{align}
   \gamma = {\left({\beta^2\over \beta_H^2} - 1\right)^{-1}}, \quad \beta_H = 2\pi \sqrt{{c\pi\mu\over6}}.
   \label{temp_ratio}
\end{align}

\begin{figure}
    \centering
    \includegraphics[width = .48\textwidth]{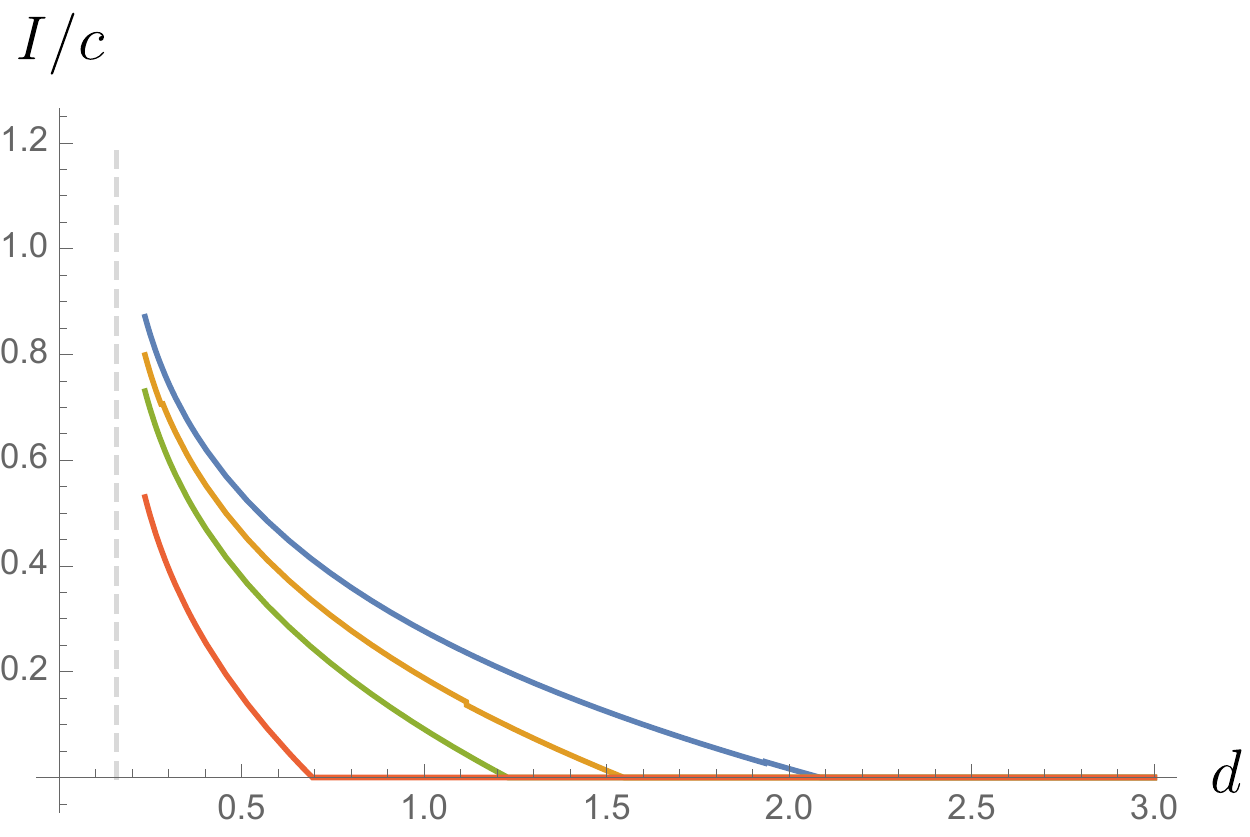}
    \includegraphics[width = .48\textwidth]{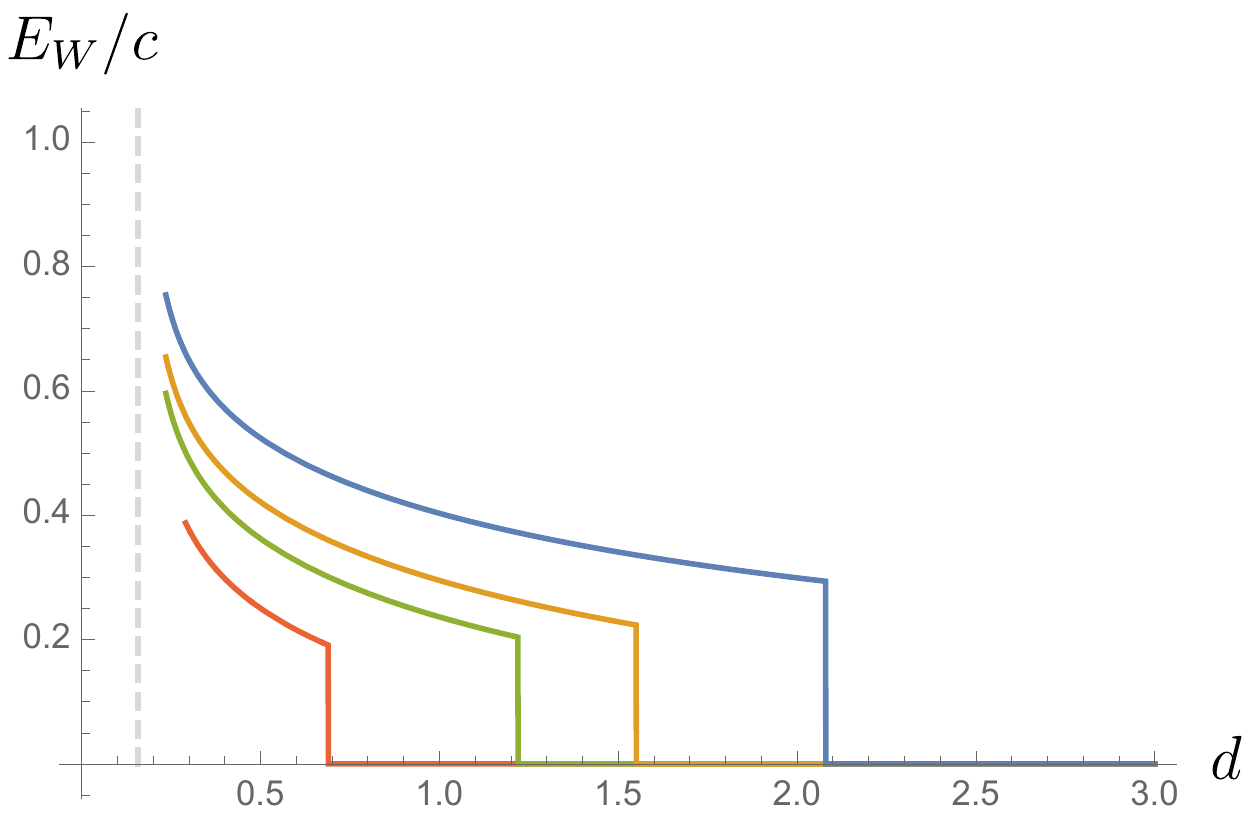}
    \caption{Left: the mutual information per central charge for disjoint intervals at finite temperature. Each subregion is of length $5$ and $\sqrt{\frac{\pi c \mu}{6}}$ is set to $\frac{1}{10}$. The temperature is varied as $\frac{\beta}{2\pi} = \{100, 3, 2,1 \}$ from blue to red respectively. 
    Right: the entanglement wedge cross section per central charge for the same parameters. We observe the divergences at $d = l_{\min}$ for both quantities. }
    \label{linear_dilaton_MI_finite_T}
\end{figure}

 The interval length $l$ in terms of the turning point of the minimal curve is given by the integral
\begin{align}
    l = {\sqrt{\alpha'}\over U_0}\int_1^{\infty} {dx\over x}\sqrt{(1 + \alpha)(1 + \alpha x)\over (x - x_{H})(x - 1)(\alpha x + \alpha + 1)}, \quad \alpha = \lambda U_0^2, \quad U_0 = e^{\phi_0},\quad x_H \leq 1,
\end{align}
which solves to 
\begin{align}
    {l\over l_{\rm min}} = {\pi\over 4}\sqrt{\alpha\over \gamma + \alpha + 1}\left[\left(1 + {1\over \gamma}\right)F(\nu, q) +\left({1\over \alpha} - {1\over \gamma}\right) \Pi \left(\nu, \left( {\gamma\over\gamma + \alpha + 
    1}\right) \left(2 + {1\over \alpha}\right), q\right)\right],
    \label{length_ftemp}
\end{align}
where
\begin{align}
    \sin\nu = \sqrt{\gamma + \alpha + 1\over 2\alpha + 1}, \quad q = \sqrt{(\gamma + 1)(2\alpha + 1)\over( \alpha  + 1)(\gamma +\alpha + 1)}.
\end{align}

We can invert the equation for the interval length to write $\alpha$ in terms of $l, \mu, c, \beta$. We then can use this to write the entropy in terms of only CFT data. The interval length at leading order in the coupling is
\begin{equation}
    {l\over {\sqrt{{c\pi\mu\over 6}}}} = {1\over \sqrt{\gamma}}\left(2{\rm arctanh}\sqrt{\gamma\over \alpha} + {\cal O}\left(\alpha^2\right)\right),\label{length_first}
    \end{equation}
To leading order in the coupling, the entropy then becomes
\begin{align}
    S_C = {c\over 6}\left[{2c\pi \mu\over 3\epsilon_{UV}^2} + \left(2 + \gamma\right)\log\left({2\beta_H\over \pi \epsilon_{UV}}\sqrt{\gamma}\sinh{{\pi l\sqrt{\gamma}\over \beta_H}} \right)\right] + {\cal O}(\mu^2).
\end{align}
At $\mu = 0$, this reduces upon using \eqref{length_first} to \eqref{univ_thermal_ent}.

The mutual information at finite temperature is obtained using the entropy \eqref{von_entropy} in \eqref{mutual_info}. We find that the mutual information is UV cutoff independent. We plot the mutual information for disjoint intervals in Fig.~\ref{linear_dilaton_MI_finite_T}. Again, there is a divergence in the mutual information at short distances when $d = l_{\min}$, the same place where the zero temperature mutual information result diverges. Heating up the system causes the information shared between the disjoint regions to monotonically decrease.

While we do not consider the disconnected regions to contribute to the von Neumann entropy, it does contribute to the entanglement wedge cross section of a single interval. The contributions from the disconnected regions are
\begin{align}
    S_D = 2\cdot {\sqrt{k\alpha'}\over 4{\rm G}_{\rm N}^{(3)}}\int_{1}^{x^H_\infty}dx{\alpha_H x + 1\over \sqrt{x(x - 1)}}, \quad \alpha_H = \lambda U_H^2, \quad x^H_\infty = {U_\infty^2\over U_H^2},
\end{align}
giving
\begin{align}
    S_D = {c\over 6}\left[{4c\pi \mu\over 3L_\Lambda^2} + (2 + \gamma)\log\left( {8c\pi \mu\over 3\gamma L_\Lambda^2} \right)\right].
\end{align}
The entanglement wedge cross section at finite temperature for an interval of length $l$ is then given by
\begin{equation}
    E_{W}(\mu, l,\gamma) = {\rm min}\left(S_D(\mu, \gamma), S_C(\mu, l, \gamma)\right).
\end{equation}






\subsubsection{Reflected entropy}

In this subsection, we compute the entanglement wedge cross section for intervals of equal length $l$ and separation distance $d$. It is given by \eqref{EW_deformed}
\begin{align}
   E_W  = \frac{c}{12}\left[\log\left({\alpha_{\rm +}\over \alpha_{\rm -}}\right) + \chi\left(\alpha_{\rm +} - \alpha_{\rm -}\right)\right] .
   \label{EW_ftemp_def}
\end{align}
Here, $\alpha_+$ and $\alpha_-$ are related to $l$ and $d$ via \eqref{length_ftemp}
\begin{align}
    {2l + d\over l_{\rm min}} &= {\pi\over 4}\sqrt{\alpha_{\rm -}\over \gamma + \alpha_{\rm -} + 1}\left[\left(1 + {1\over \gamma}\right)F(\nu, q) +\left({1\over \alpha_{\rm -}} - {1\over \gamma}\right) \Pi \left(\nu, \left( {\gamma\over\gamma + \alpha_{\rm -} + 
    1}\right) \left(2 + {1\over \alpha_{\rm -}}\right), q\right)\right],
    \nonumber
\\
    { d\over l_{\rm min}} &= {\pi\over 4}\sqrt{\alpha_{\rm +}\over \gamma + \alpha_{\rm +} + 1}\left[\left(1 + {1\over \gamma}\right)F(\nu, q) +\left({1\over \alpha_{\rm +}} - {1\over \gamma}\right) \Pi \left(\nu, \left( {\gamma\over\gamma + \alpha_{\rm +} + 
    1}\right) \left(2 + {1\over \alpha_{\rm +}}\right), q\right)\right].
\end{align}
In what follows, we study the first few leading terms in \eqref{EW_ftemp_def} in perturbation theory. At $\lambda = 0$, we find using \eqref{length_first}
\begin{equation}
    E_{W} =  {c\over 6}\log\left[{\coth{{\pi d\over \beta}}\over \coth{\pi (2l + d)\over \beta}}\right].
\end{equation}
The ${\cal O}(\lambda)$ correction comes from the non-logarithmic term in \eqref{EW_ftemp_def}, and including this we get (upon using \eqref{length_first})
\begin{align}
    E_{W} &= {c\over 6}\log\left[{\coth{{\pi d\over \beta}}\over \coth{\pi (2l + d)\over \beta}}\right] + {c\over 6}\cdot {\lambda \over 2}\cdot U_H^2\left({\rm coth}^2\left({d\over 2\sqrt{\alpha'}}U_H\right) - {\rm coth}^2\left({2l + d\over 2\sqrt{\alpha'}}U_H\right)\right) + {\cal O}(\lambda^2)
    \nonumber
    \\
    &= {c\over 6}\log\left[{\coth{{\pi d\over \beta}}\over \coth{\pi (2l + d)\over \beta}}\right] + {c\pi^2\alpha'\lambda\over 48\beta^2}\left({\rm coth}^2\left({\pi d\over 4\beta}\right) - {\rm coth}^2\left( {\pi(2l + d)\over 4\beta}\right)\right) + {\cal O}(\lambda^2).
\end{align}
Expanding the ${\cal O}(\lambda)$ correction in powers of the inverse temperature $\beta$ gives
\begin{align}
   \lambda {\partial E_{W}\over \partial \lambda} &= {4cl(d+l)\alpha'\lambda\over 3d^2(d + 2l)^2} -{4cl(d + l)\pi^4\alpha'\lambda\over 45\beta^4} + {16 cl(d+ l)(d^2 + 2dl+2l^2)\pi^6\alpha'\lambda\over 567\beta^6} + {\cal O}\left({1\over \beta^8}\right)
\nonumber
\\
    &= {2\pi c^2l(d+l)\mu \over 9d^2(d + 2l)^2} -{2 c^2 l(d + l)\pi^5\mu\over 135\beta^4} + {8 c^2 l(d+ l)(d^2 + 2dl+2l^2)\pi^7\mu\over 1701\beta^6} + {\cal O}\left({1\over \beta^8}\right).
    \label{finite_temp_lin_dil_EW_pert}
\end{align}
We note that the order $\mu$ small temperature leading correction is negative. We plot the entanglement wedge cross section in Fig.~\ref{linear_dilaton_MI_finite_T}. Both the finite distance divergence as $d$ approaches the non--locality scale $l_{\rm min}$ and the monotonic decrease with temperature is clear.

\subsection{Comments on the split property}

We briefly comment on some interesting consequences of the computations in this section. We have found the mutual information and reflected entropy to generically diverge when the distance between intervals approaches $\beta_H/4$. This means that the split property must have failed i.e.~there does not exist a type I factor that splits $\mathfrak{A}(A)$ and $\mathfrak{A}(\bar{B})$. In the geometric regularization scheme, we are then unable to take the $L_A/\epsilon \rightarrow 0$ limit. This limits our ability to extract the relevant physical constants that serve as $c$-functions. It is an important question how to define $c$-functions for theories like these that are non-local at short distances along the renormalization group flow towards the UV.

An additional curiosity is the factor of $4$ in $l_{\min}$. In theories with Hagedorn divergences, the minimum distance for a split is $\beta_H$ \cite{Haag:1992hx}, but we do not see the divergence in the reflected entropy until $d = \beta_H/4$. 
For $T\bar T$-deformed theories, the Hagedorn density of states only starts to dominate at some energy scale (which depends quadratically on $\beta_H$), thus $\beta_H$ only acts as an upper bound for the minimum distance needed for a split. This factor of $4$ is therefore quite nontrivial.
We note that the torus partition function itself diverges precisely at $\beta = \beta_H$.

\section{Double-trace \texorpdfstring{$T\bar T$}{TEXT} and cutoff AdS}
\label{sec_cutoffAdS}

We now compute extremal surfaces in the cutoff AdS geometry proposed to be holographically dual to the double-trace $T\bar{T}$ deformation \cite{2016arXiv161103470M}. Importantly, this prescription uses the opposite sign of the deformation parameter\footnote{In this section, we will always take $\mu < 0$ so the deformation of the Lagrangian is $\mathcal{L} \rightarrow \mathcal{L} - \mu \int T \bar T$.} such that the spectrum always contains complex energies.

Starting with the metric for the BTZ black hole
\begin{align}
    ds^2 = \frac{r^2-r_H^2}{l_{AdS}^2}dt^2+\frac{l_{AdS}^2}{r^2-r_H^2}dr^2+r^2dx^2,
    \label{metric}
\end{align}
where the asymptotic boundary is at $r\rightarrow \infty$ and the horizon is at $r_H$, the deformation corresponds to a finite radial cutoff with Dirichlet boundary conditions with
\begin{align}
    r_c^2 = -\frac{6l_{AdS}^4}{\mu \pi c}.
\end{align}
The horizon radius is related to the temperature as
\begin{align}
    r_H = \frac{2\pi }{\beta}.
\end{align}
The metric on the finite cutoff boundary is
\begin{align}
    ds^2 = \frac{r_c^2-r_H^2}{l_{AdS}^2}dt^2+r_c^2dx^2.
\end{align}
Thus, we must rescale the metric such that $t$ is a physical time on this surface
\begin{align}
    ds^2 \rightarrow  dt^2+\frac{l_{AdS}^2r_c^2}{r_c^2-r_H^2}dx^2.
\end{align}
Note that this subtlety is absent at zero temperature ($r_H\rightarrow 0$).

\subsection{Vacuum}

\subsubsection{Mutual information}

At zero temperature, the von Neumann entropy of an interval of length $l$ can straightforwardly be computed
\begin{align}
    S(l, \mu) = \frac{c}{6}\cosh^{-1}\left[1 - \frac{3 l^2}{\mu \pi c} \right] .
\end{align}
The mutual information between disjoint intervals is simply
\begin{align}
    I = \max\left[S(l_A, \mu)+S(l_B, \mu) -S(l_A+l_B+d, \mu)-S(d, \mu) ,0\right].
    \label{MI_hol_def}
\end{align}
It is interesting to consider the adjacent intervals limit ($d\rightarrow 0$) of the mutual information. Interestingly, the mutual information in this limit is UV finite for any value of $\mu$. At leading order in the deformation parameter, we have
\begin{align}
    I = \frac{c}{3}\log\left[\frac{l_A l_B}{(l_A+l_B)\sqrt{-\frac{\mu \pi c}{6}}}\right].
\end{align}
This is identical to the universal value of the mutual information for adjacent intervals in a conformal field theory \cite{1994NuPhB.424..443H} if one identifies $\sqrt{-\frac{\mu \pi c}{6}}$ with the UV cutoff. Thus, the deformation provides a natural cutoff for this sign of the deformation.

\begin{figure}
    \centering
    \includegraphics[width = .48\textwidth]{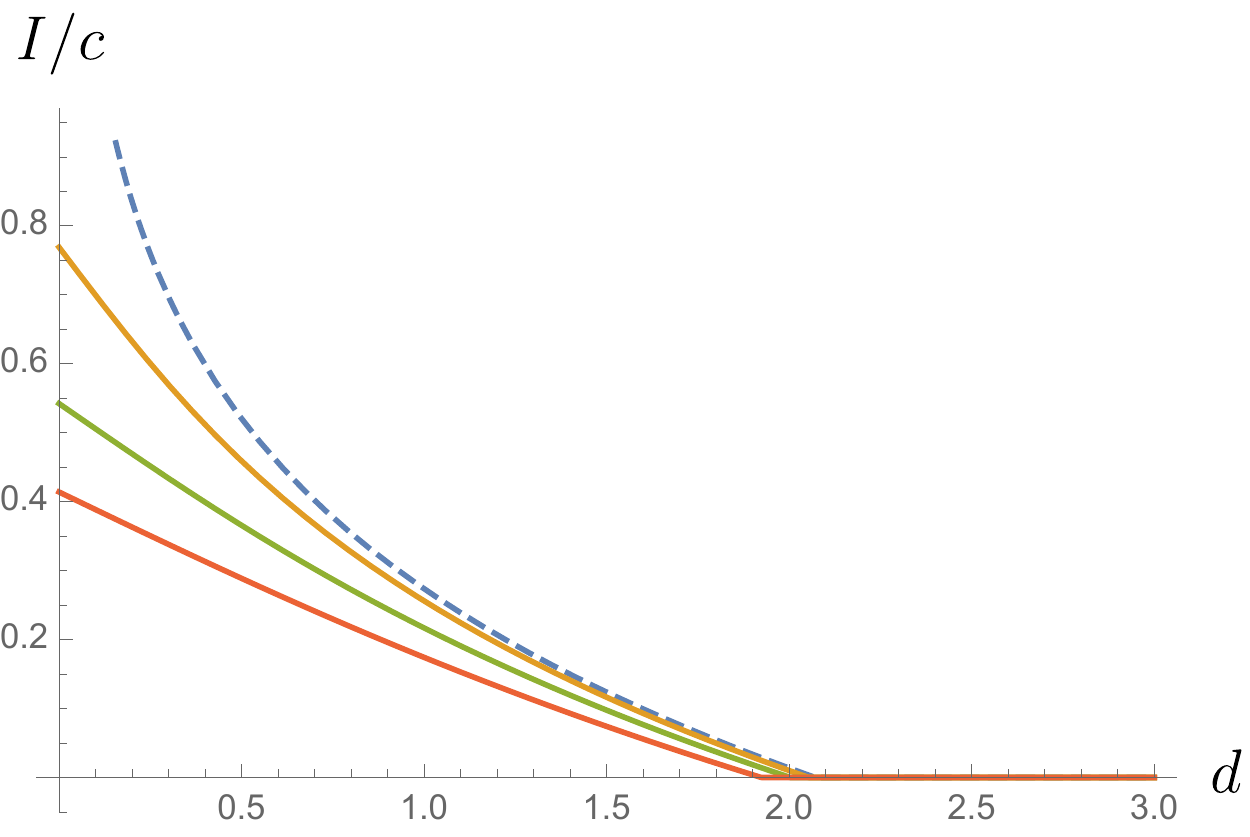}
    \includegraphics[width = .48\textwidth]{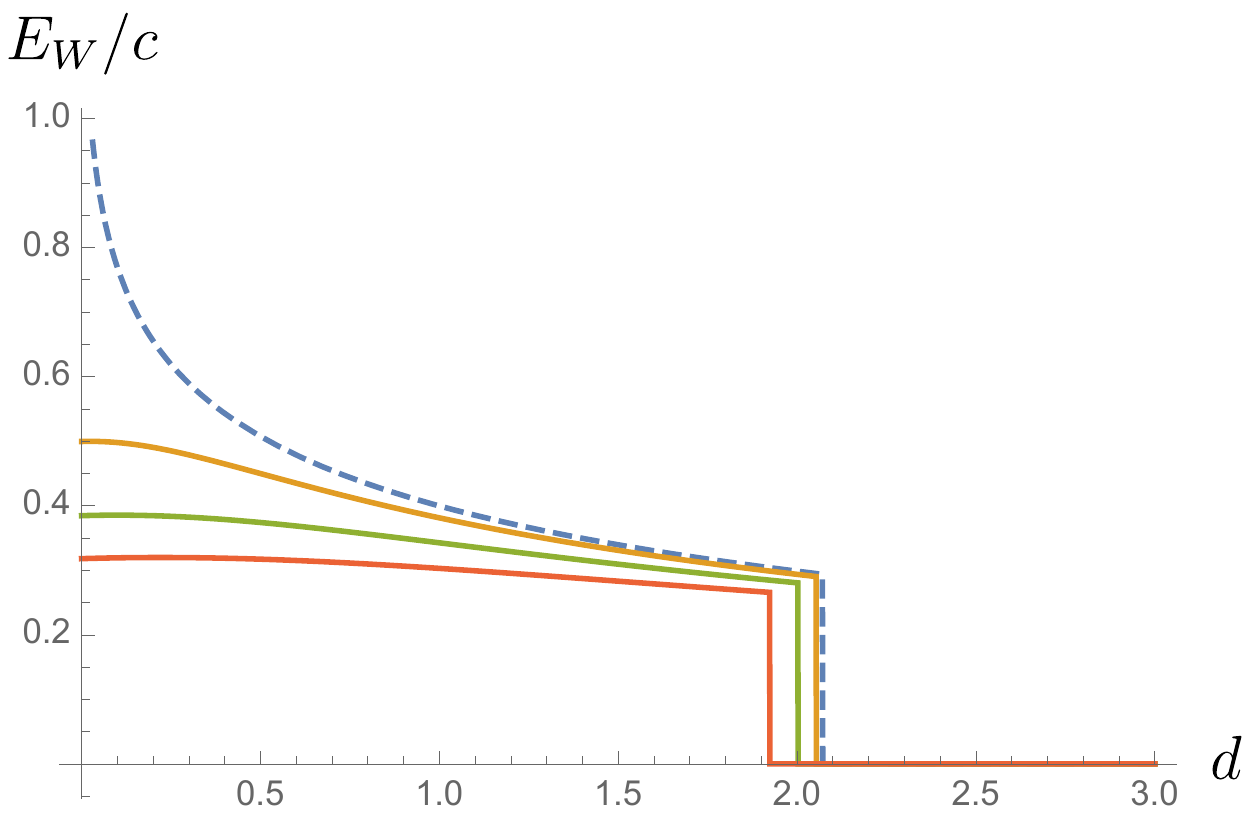}
    \caption{The mutual information (left) and entanglement wedge cross section (right) for $z_c = \{0,\frac{1}{4},\frac{1}{2},\frac{3}{4} \}$ in descending order. When the deformation parameter is finite (solid lines), $E_W$ and $I$ are manifestly finite even as the intervals become adjacent. We use symmetric intervals of length $l =5$.}
    \label{cutoff_EW_varyd}
\end{figure}

We stress that this result is extremely different from that of Section \ref{sec_lin_dilaton} where the value $\sqrt{\frac{\mu \pi c}{6}}\equiv l_{min}$ determined the \textit{finite} distance between the intervals where the mutual information \textit{diverged}. Clearly, the correlation structure is drastically changed in opposite ways given the sign of the deformation. The results are plotted in Fig.~\ref{cutoff_EW_varyd}.

\subsubsection{Reflected entropy}

We now investigate how this UV regulation emerges for the reflected entropy by returning to the entanglement wedge cross section. As previously noted, at zero temperature and $\mu = 0$, the area of the entanglement wedge cross section is a simple function of the conformally invariant cross-ratio \cite{2018NatPh..14..573U}
\begin{align}
    E_W = \frac{c}{6} \log \left(\frac{1+\sqrt{x}}{1-\sqrt{x}} \right), \quad x = \frac{\tilde{x}_{21}\tilde{x}_{43}}{\tilde{x}_{42}\tilde{x}_{31}},
    \label{undeformed_EW}
\end{align}
where the coordinates with tildes correspond to those on the asymptotic boundary.

\begin{figure}
    \centering
    \includegraphics[width = .5\textwidth]{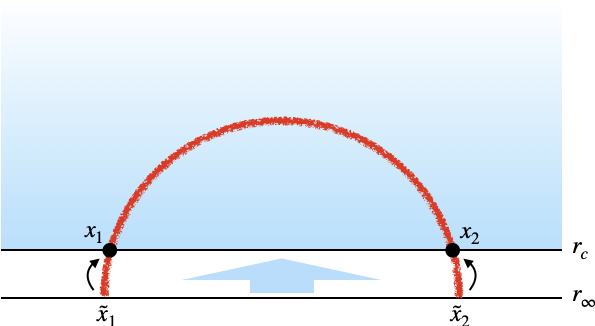}
    \caption{In the cutoff AdS approach, the standard Dirichlet boundary conditions of holography are moved from the conformal boundary at $r_{\infty}$ to a hard Dirichlet boundary at $r_c$. Because the geometry within the new boundary remains the same, so do the geodesics. In the figure, we show how the entangling surface points of the interval are mapped from the asymptotic boundary to the finite boundary.}
    \label{radial_cutoff_map}
\end{figure}

In the majority of parameter space, the entanglement wedge cross section has the same area as the undeformed theory because it lies in the IR part of the geometry which is entirely unchanged in the cutoff AdS story. Though the area of the geometric object remains unchanged, the dependence of the cross-ratio on the boundary positions will flow. In particular, the cross-ratio will no longer be conformally invariant. We can map the points from the asymptotic boundary to the finite cutoff boundary along the geodesics (see Fig.~\ref{radial_cutoff_map}), which are semi-circles at zero temperature when we work in Poincar\'e coordinates ($r\rightarrow l_{AdS}^2/z$)
\begin{align}
    r = \frac{ \sqrt{l^2 - 4x^2}}{2}.
\end{align}
This leads to the following mapping of boundary coordinates
\begin{align}
    \tilde{x}_1 = \frac{1}{2}\left( x_1 + x_4 - \sqrt{(x_1-x_4)^2 + 4z_c^2}  \right),
    \\
    \tilde{x}_2 = \frac{1}{2}\left( x_2 + x_3 - \sqrt{(x_2-x_3)^2 + 4z_c^2}  \right),
    \\
    \tilde{x}_3 = \frac{1}{2}\left( x_2 + x_3 + \sqrt{(x_2-x_3)^2 + 4z_c^2}  \right),
    \\
    \tilde{x}_4 = \frac{1}{2}\left( x_1 + x_4 + \sqrt{(x_1-x_4)^2 + 4z_c^2}  \right).
\end{align}
Thus, in terms of the cutoff coordinates, the cross-ratio is
\begin{align}
x = \frac{-\sqrt{(x_1-x_4)^2+4 z_c^2} \sqrt{(x_2-x_3)^2+4
   z_c^2}+x_1 (x_2+x_3-2 x_4)+x_4
   (x_2+x_3)-2 x_2 x_3+4
   z_c^2}{\sqrt{(x_1-x_4)^2+4 z_c^2}
   \sqrt{(x_2-x_3)^2+4 z_c^2}+x_1 (x_2+x_3-2
   x_4)+x_4 (x_2+x_3)-2 x_2 x_3+4 z_c^2}.
   \label{deformed_cross}
\end{align}
Inserting \eqref{deformed_cross} into \eqref{undeformed_EW}, we arrive at the full non-perturbative area of the entanglement wedge cross section. There is a correction at leading order in $\mu$. For e.g.~equal length intervals of length $l$ and distance $d$, we find
\begin{align}
     E_W = \frac{c}{6}\log \left[1+ \frac{2l}{d}\right] +\frac{2c^2 l(d+l)\pi \mu}{9d^2(d+2l)^2} + O(\mu^2).
\end{align}
Interestingly, this leading order correction is identical to the one found for the linear dilaton geometry \eqref{zero_temp_lin_dil_EW_pert}. However, we again stress that we take $\mu < 0$ such that this \textit{decreases} the correlations, the opposite effect found from the linear dilaton analysis. The sign of the change in correlations will crucially depend on how one ``flows up the RG.''

This correction, of course, is only valid when we are within the connected regime of the entanglement wedge ($I>0$). There is a phase transition of the entanglement wedge to the disconnected regime which depends on the deformation parameter.
The phase transition occurs when $S_{[x_1, x_2]}+S_{[x_3, x_4]} =S_{[x_1, x_4]}+S_{[x_2, x_3]} $. In the disconnected regime, $E_W = 0$, so $\Delta E_W= 0$. We plot the corresponding $E_W$ and mutual information in Fig.~\ref{cutoff_EW_varyd}.

Let us now look at the adjacent intervals limit. We find $E_W$ to be UV finite as $d\rightarrow 0$ for any finite value of $\mu$. In particular, the leading order correction (in $\mu$) is
\begin{align}
    E_W = \frac{c}{6}\log\left[\frac{2l_A l_B}{(l_A+l_B)\sqrt{-\frac{\mu \pi c}{6}}}\right].
\end{align}
Analogous to the mutual information, this is equivalent to the adjacent intervals limit of (half of the) reflected entropy in a conformal field theory once identifying $\sqrt{-\frac{\mu \pi c}{6}}$ with the UV cutoff. The results are plotted in Fig.~\ref{cutoff_EW_varyd}.

\subsection{Finite temperature}

We progress to finite temperature where it will be easier to compare with boundary computations of the following section. For finite temperature calculations, we simply set $r_H > 0$ in \eqref{metric}.

\subsubsection{Mutual information}

The von Neumann entropy for an interval can be found at finite temperature from the Ryu-Takayanagi formula
\begin{align}
    S = \frac{c}{6}\cosh^{-1}\left[1 + 2 \left(\frac{r_c}{r_H}\right)^2\sinh^2\left[\frac{r_H{l\sqrt{r_c^2-r_H^2}}}{2 r_c} \right] \right].
\end{align}
This is UV finite and transitions from logarithmic growth for $l$ smaller than the thermal length to linear growth for large $l$. The linear growth signals that it operates as an extensive thermodynamic entropy. The mutual information is again determined by \eqref{MI_hol_def}. To leading order in the deformation parameter,
\begin{align}
    I &= \frac{c}{6} \log \left[\frac{\sinh^2\left(\frac{l\pi}{\beta}\right)}{\sinh\left(\frac{d\pi}{\beta}\right) \sinh\left(\frac{(2l+d)\pi}{\beta}\right)}\right] - \mu \frac{c \pi^3}{18 \beta^3}\Big( d\pi \coth \left(\frac{d\pi}{\beta} \right) - 2l\pi \coth \left(\frac{l\pi}{\beta} \right) 
    \nonumber
    \\
    &+ (d+2l)\pi \coth \left(\frac{(2l+d)\pi}{\beta} \right) - \beta \left(\csch \left(\frac{d\pi}{\beta}\right)+\csch \left(\frac{(2l+d)\pi}{\beta}\right)-2\csch \left(\frac{l\pi}{\beta}\right)\right)\Big).
\end{align}

\subsubsection{Reflected entropy}
\begin{figure}
    \centering
    \includegraphics[width = .5\textwidth]{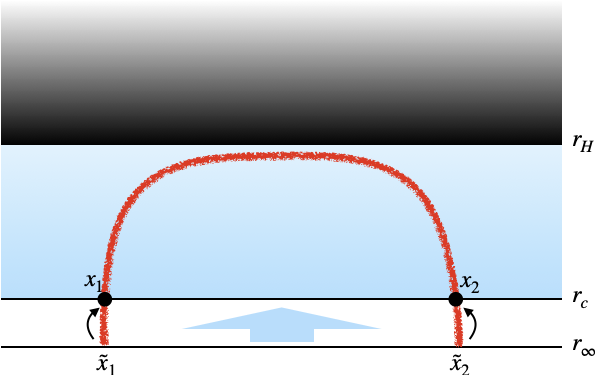}
    \caption{The analogue of the mapping in Fig.~\ref{radial_cutoff_map}, this time with the additional scale, $r_H$, representing the black hole horizon radius.}
    \label{radial_cutoff_map_BH}
\end{figure}

To solve for the entanglement wedge cross section, we again map the coordinates on the asymptotic boundary to the cutoff surface (Fig.~\ref{radial_cutoff_map_BH}). For this, we need to solve the geodesic equation for a boundary anchored curve giving a parametrization of the Ryu-Takayanagi surface 
\begin{align}
    r(x) = r_H\left( 1 - \frac{\cosh^2 (r_H x)}{\cosh^2 \left(\frac{r_H l}{2}\right)}\right)^{-1/2}.
\end{align}
The turning point is at
\begin{align}
    r_* = \frac{r_H}{\tanh \left( r_H \frac{\tilde{l}}{2}\right)}.
\end{align}
The length of the interval on the cutoff surface in terms of the length at the asymptotic boundary is
\begin{align}
    l = \frac{2 r_c\cosh^{-1}\left[\frac{\cosh\left(r_H \frac{\tilde{l}}{2}\right)\sqrt{r_c^2-r_H^2}}{r_c} \right]}{r_H\sqrt{r_c^2 - r_H^2}},
\end{align}
or inverted
\begin{align}
    \tilde{l} = \frac{2 \cosh^{-1}\left[\frac{r_c \cosh \left(\frac{ l r_H \sqrt{rc^2 - r_H^2} }{2r_c}\right)}{\sqrt{r_c^2 -r_H^2}} \right]}{r_H},
\end{align}
where we have made sure to be careful in rescaling the physical length of the interval at the finite cutoff. 

\begin{figure}
    \centering
    \includegraphics[width = .48\textwidth]{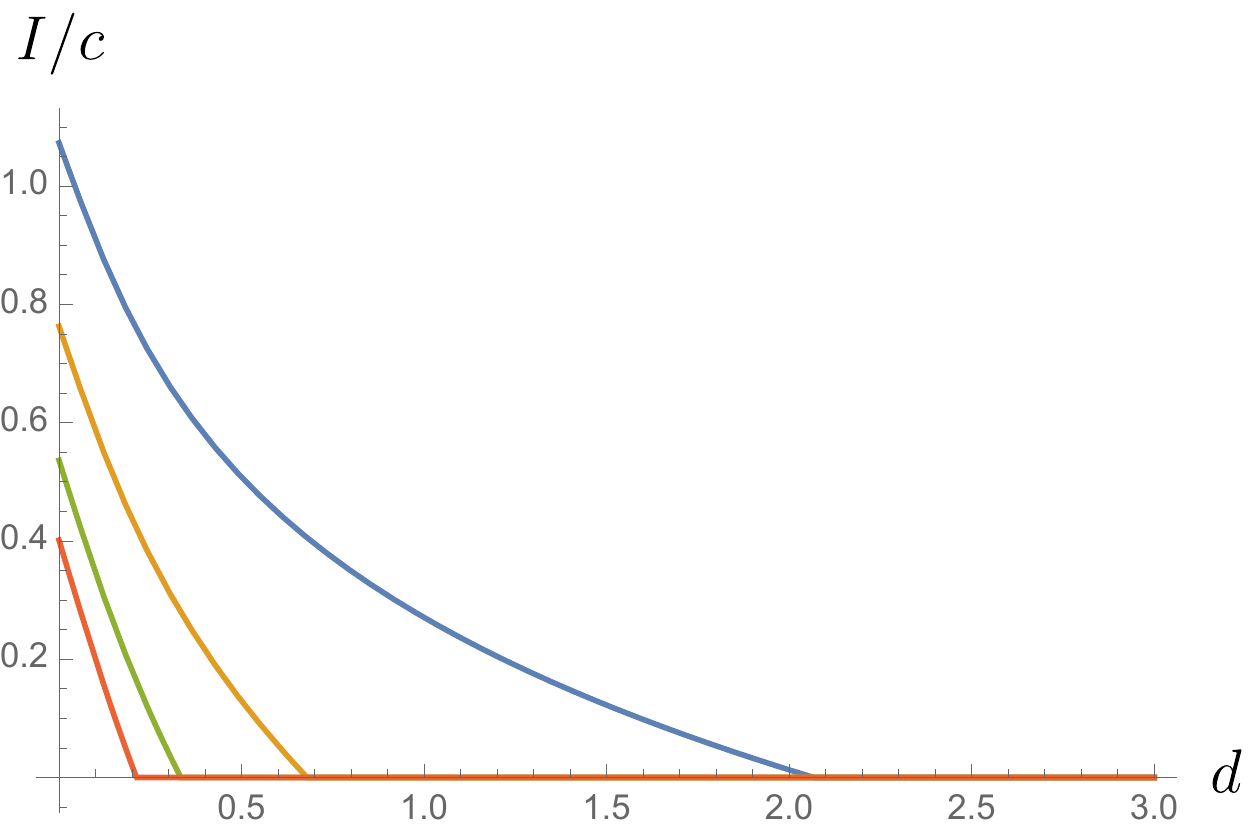}
    \includegraphics[width = .48\textwidth]{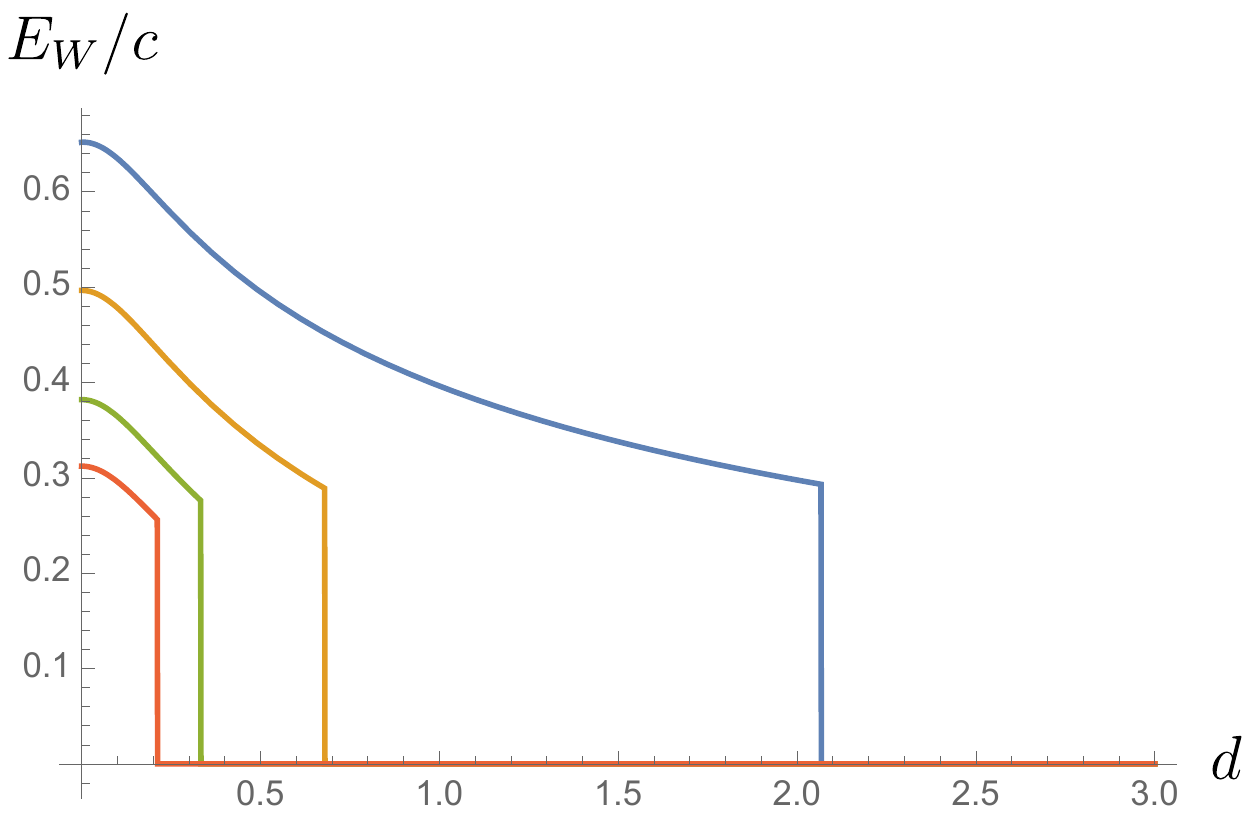}
    \caption{Left: The mutual information per unit central charge for disjoint intervals of equal length ($5$). We set the radial cutoff to $r_c = 10$ and vary the temperature $r_H = \{10^{-2}, 1,2,3 \}$ from blue to red respectively. Right: The same configurations, but for the entanglement wedge cross section .}
    \label{finiteT_cutoffAdS}
\end{figure}

For simplicity, we consider disjoint intervals of equal length, so that the entanglement wedge cross section is purely radial
\begin{align}
    E_W = \frac{c}{6} \int_{r_*(2l+d)}^{r_*(d)} \frac{dr}{\sqrt{r^2-r_H^2}} = \frac{c}{6}\tanh^{-1}\left[\frac{r}{\sqrt{r^2 - r_H^2}} \right] \Big|_{r_*(2l+d)}^{r_*(d)}.
\end{align}
Evaluating this is straightforward but gives a complicated, unenlightening expression for the non-perturbative correction to $E_W$. Instead, we consider the first order correction (linear in $\mu$)
\begin{align}
    E_{W} &= \frac{c}{6}\log\left[{\coth{{\pi d\over \beta}}\over \coth{\pi (2l + d)\over \beta}}\right]
    \nonumber
    \\&-\frac{\pi ^3 c^2 \mu  \left(\left(\pi  d-\beta  \coth \left(\frac{\pi  d}{\beta }\right)\right)
   \csch\left(\frac{\pi  d}{\beta }\right)+\left(\beta  \coth \left(\frac{\pi  (d+2 l)}{\beta }\right)+\pi 
   (-d-2 l)\right) \csch\left(\frac{\pi  (d+2 l)}{\beta }\right)\right)}{18 \beta ^3} +O(\mu^2).
\end{align}
In the low temperature expansion (taking $\mu\rightarrow 0$ before $\beta \rightarrow \infty$), this gives
\begin{align}
    E_{W} &= \frac{c}{6}\log\left[{\coth{{\pi d\over \beta}}\over \coth{\pi (2l + d)\over \beta}}\right]
    \nonumber
    \\
    &+{2\pi c^2\mu l (l + d)\over 9 d^2 (d + 2l)^2} - \frac{13c^2l(d+l)\pi^5\mu}{540 \beta^4} + \frac{139c^2\mu\pi^7l(d+l)(d^2+2dl+2l^2)}{34020\beta^6} + O(\beta^{-8}).
    \label{finite_temp_cutoff_ADS_EW_pert}
\end{align}
This has identical functional form to the linear dilaton results \eqref{finite_temp_lin_dil_EW_pert}, though the numerical coefficients, after the zero temperature correction, disagree. We plot the non-perturbative results in Fig.~\ref{finiteT_cutoffAdS} where it is clear that heating up the system destroys the quantum correlations between the subsystems. Again, $E_W$ is UV finite for all $d$. 

\section{Conformal perturbation theory}
\label{sec_CFT}

It is important to check whether the predictions from holography from the previous section agree with (perturbed) CFT computations. In order to non-perturbatively understand how the entanglement structure changes under the flow induced by the irrelevant deformations from the field theory side, one must understand how the partition function on arbitrary genus Riemann surfaces flows. While there has been significant progress for sphere and torus partition functions \cite{2016JHEP...10..112C,2018JHEP...09..158D,2018JHEP...08..106D,2018PhRvL.121m1602D,2019arXiv190911118H,2019arXiv190707221H,2019arXiv190500051C,2019JHEP...01..085A,2019JHEP...01..086A,2019arXiv191204686I,2019JHEP...05..112C}, little is known about more generic higher genus partition functions. This is why, thus far, the only non-perturbative results for entanglement measures under $T\bar{T}$ flows is for the very special case of von Neumann entropy where the replica Riemann surface is genus zero \cite{2018PhRvL.121m1602D,2020JHEP...04..152L}. The one exception to this is for 2D Yang-Mills theory because of its near topological behavior \cite{2019arXiv191204686I}. 

We consider the partition function of the $T\bar{T}$ deformed theory from the path integral formalism
\begin{align}
    \mathcal{Z}_{n} = \int \left[\mathcal{D}\phi_n\right]e^{-S_{CFT}+\mu \int_{\mathcal{M}_{n}} d^dxT\bar{T} },
\end{align}
where $\mathcal{M}_{n}$ is the replica manifold of $n$ connected copies of the theory which generically has highly non-trivial topology. 
We work perturbatively so that we can expand in the coupling as
\begin{align}
    \mathcal{Z}_{n} &= \int \left[\mathcal{D}\phi_n\right]e^{-S_{CFT}}\left(1+\mu \int_{\mathcal{M}_{n}}d^dx(T\bar{T} )+\mathcal{O}(\mu^2)\right),\\
    &= \int \left[\mathcal{D}\phi_n\right] e^{-S_{CFT}}\left(1+\mu \int_{\mathcal{M}_{n}}d^dx\langle T\bar{T}\rangle_{\mathcal{M}_{n}} +\mathcal{O}(\mu^2)\right),
\end{align}
where we have used the definition
\begin{align}
    \langle T \bar T\rangle_{\mathcal{M}_n} \equiv \frac{\int \left[\mathcal{D}\phi_n\right]e^{-S_{CFT}} (T \bar T)}{\int \left[\mathcal{D}\phi_n\right]e^{-S_{CFT}}}.
\end{align}
To leading order in $\mu$, the change in the logarithm of the partition function due to the deformation is
\begin{align}
    \delta \log \mathcal{Z}_{n} = \mu
    \int_{\mathcal{M}_{n}}\langle T\bar{T}\rangle_{\mathcal{M}_{n}}.
\end{align}
Generalizing to include the other irrelevant deformations, we have
\begin{align}
    \delta \log \mathcal{Z}_{n} = \mu
    \int_{\mathcal{M}_{n}}\langle T\bar{T}\rangle_{\mathcal{M}_{n}}+
    \mu_+
    \int_{\mathcal{M}_{n}}\langle J\bar{T}\rangle_{\mathcal{M}_{n}}+
    \mu_-
    \int_{\mathcal{M}_{n}}\langle T\bar{J}\rangle_{\mathcal{M}_{n}}.
\end{align}

\subsection{\texorpdfstring{$T\bar{T}$}{TEXT} at finite temperature}

We will focus on the reflected entropy at finite temperature because there are ambiguous contact terms in the correlation functions at zero temperature. Similar ambiguities have previously been commented on in Ref.~\cite{2018NuPhB.935..290C}. To compute the reflected entropy directly in the field theory, we must compute the path integral on an $mn$-sheeted branched cover of the CFT
\begin{align}
    S_R = \lim_{n,m\rightarrow 1}\frac{1}{1-n}\log \frac{\mathcal{Z}_{n,m}}{\left(\mathcal{Z}_{1,m}\right)^n},
\end{align}
where $m \in 2\mathbb{Z}$ is the replica index for the given purification and $n \in \mathbb{Z}$ is the standard replica index for the R\'enyi entropies. Recall that in holographic theories, the logarithmic negativity is equivalent to the following path integral when taking $n\rightarrow 1/2$ instead of $n\rightarrow 1$. The path integrals may be computed by a correlation function of twist fields in the $S_{mn}$ orbifold theory. For example, when we consider the reflected entropy between two disjoint intervals, we must compute \cite{2019arXiv190500577D}
\begin{align}
    \mathcal{Z}_{n,m} = \langle  \sigma_{g_A}(z_1) {\sigma}_{g_A^{-1}}(z_2) {\sigma}_{g_B}(z_3) \sigma_{g_B^{-1}}(z_4) \rangle,
\end{align}
where the twist fields have conformal dimensions 
\begin{align}
    h_{g^{\ }_B} = h_{g^{-1}_B}= h_{g_A^{\ }} = h_{g_A^{-1}} = \frac{cn(m^2-1)}{24m}.
    \label{SR_twist_dim}
\end{align}
In the operator product expansion of  $\sigma_{g_A^{-1} }$ and $\sigma_{g_B}$,  the leading operator is another Virasoro primary field, $\sigma_{g_A^{-1} g_B}$, with conformal dimension
\begin{align}
    h_{g^{\ }_B g_A^{-1}} = \frac{2c(n^2-1)}{24n}.
\end{align}

We will approximate these four-point functions by taking only the dominant conformal block in the conformal block decomposition. This is a valid approximation in the limit of large central charge because contributions from subdominant conformal blocks are exponentially suppressed in $c$ \cite{Zamolodchikov:1985ie,2020JHEP...01..109B}. When considering the $J\bar{T}$ and $T\bar{J}$ deformations, we, by definition, have a conserved $U(1)$ current. We are then required to take the dominant $Vir \times U(1)$ block rather than just the Virasoro conformal block. This extended conformal block was shown to factorize at large $c$ into Virasoro and $U(1)$ blocks as \cite{2015JHEP...11..200F}
\begin{align}
    \mathcal{V}_{T+J}(c,h_i,k,q_i,h_p,z) = \mathcal{V}_T\left(c-1,h_i-\frac{q_i^2}{2k},h_p,z\right) \mathcal{V}_J(k,q_i,z),
\end{align}
where $q_i$ are the $U(1)$ charges and $h_p$ is the conformal weight of the intermediate operator. $\mathcal{V}_T$ is the Virasoro block that may be evaluated in the heavy-heavy-light-light limit as
\begin{align}
    \mathcal{V}_T = (1-z)^{h_L(\alpha-1)}\left( \frac{1-(1-z)^{\alpha}}{\alpha}\right)^{h_p-2h_L} {}_2F_1\left(h_p,h_p,2h_p,1-(1-z)^{\alpha} \right),
    \nonumber
    \\
    \alpha \equiv \sqrt{1- \frac{24h_H}{c}},
    \label{HHLL}
\end{align}
and $\mathcal{V}_J$ is the contribution only from the $U(1)$ descendent states
\begin{align}
    \mathcal{V}_J = z^{-\frac{q_L^2}{k}}(1-z)^{\frac{q_Hq_L}{k}}.
    \label{U1_cont_block}
\end{align}
Here, $k$ is the level of the $U(1)$ current algebra
\begin{align}
    [J_n, J_m] = n k\delta_ {n+m,0},
\end{align}
and in the linear dilaton story is related to the AdS radius, $l_{AdS}$, as
\begin{align}
    k = \frac{l_{AdS}^2}{l_s^2}.
\end{align}
Thus, in the semiclassical limit in the bulk, the level will be large.

We now specify to two disjoint intervals where the correlation functions on $\mathcal{M}_{n,m}$ may be computed by correlation functions of twist fields on $\mathcal{M}$
\begin{align}
    \langle T\bar{T}\rangle_{\mathcal{M}_{n,m}} =  \frac{1}{nm}\langle  T(w)\bar{T}(\bar{w}) \sigma_{g_A}(z_1) {\sigma}_{g_A^{-1}}(z_2) {\sigma}_{g_B}(z_3) \sigma_{g_B^{-1}}(z_4) \rangle_{\mathcal{M}},
\end{align}
where the stress tensors on the right hand side live in the orbifold theory, leading to the prefactor. A similar expression holds for the other two terms. The strategy is to apply the Ward identities to evaluate these correlation functions and then integrate over the thermal cylinder. The calculation is technical and involves long formulas and subtleties, so we relegate some details to Appendix \ref{app_details}. 

After evaluating, we find the change in the reflected entropy to be
\begin{dmath}
\label{maintext_pert_EW}
   \Delta S_R = \frac{\pi ^4 c^2 \mu  e^{\frac{\pi  d}{\beta }} \left(\coth \left(\frac{\pi  d}{\beta }\right)-1\right) \left(2 l \left(e^{\frac{2 \pi  d}{\beta }}-1\right) e^{\frac{2 \pi  l}{\beta
   }}-d \left(e^{\frac{2 \pi  l}{\beta }}-1\right) \left(e^{\frac{2 \pi  (d+l)}{\beta }}+1\right)\right) \left(\coth \left(\frac{\pi  (d+2 l)}{\beta }\right)-1\right)}{18 \beta ^3}.
\end{dmath}
The zero temperature ($\beta\rightarrow \infty$) limit is
\begin{align}
    \lim_{\beta\rightarrow \infty} \Delta S_R = 0.
\end{align}
This, however, is not sensitive to the contact terms discussed earlier. Similarly, the $\beta \rightarrow \infty$ limit of the von Neumann entropy is zero even though there are finite corrections at zero temperature due to contact terms. The leading low temperature correction comes at fourth order
\begin{align}
    S_R &= \frac{c}{3} \log\left[{\coth{{\pi d\over \beta}}\over \coth{\pi (2l + d)\over \beta}}\right]
    -  \frac{2c^2l(d+l)\pi^5\mu}{27 \beta^4} + \frac{7c^2\mu\pi^7l(d+l)(d^2+2dl+2l^2)}{405\beta^6} + O(\beta^{-8}).
\end{align}


Besides the zero temperature term which may arise from a contact term, we note that it is interesting that this has precisely the same form as both holographic calculations \eqref{finite_temp_lin_dil_EW_pert} and \eqref{finite_temp_cutoff_ADS_EW_pert}. However, the numerical coefficients are different. This suggests that the correlation structures of the $T\bar T$ deformed CFT and the field theory dual to cutoff AdS geometry are very similar, but not quite identical. Thus, this suggests that modifications, perhaps similar to Refs.~\cite{Kraus:2018xrn,2019arXiv190611251G}, must be made to have a precise agreement for the two theories\footnote{Another logical possibility is that non-perturbative corrections restore consistency.}. We reiterate that the cutoff AdS results assumed the validity of holographic duality between reflected entropy and the entanglement wedge cross-section in generic spacetimes, an assumption that is on solid footing but not rigorously derived \cite{2019arXiv190500577D,Murdia:2019fax}. In order to claim modifications must be made for the linear dilaton proposal, we would have had to compute the single-trace correction in the CFT.

\subsection{\texorpdfstring{$J\bar{T}$}{TEXT} at finite temperature}
We can also apply the Ward identities for the currents for the $J \bar{T}$ correction
\begin{dmath}
    \int_{\mathcal{M}^{(n,m)}} \langle J \bar{T}\rangle_{\mathcal{M}^{(n,m)}} = \frac{1}{\langle \dots\rangle}_{\mathbb{C}} \int_{\mathcal{M} }\frac{1}{nm} \left[\left(\frac{2\pi z}{\beta} \right)\sum_{j=1}^k\left(\frac{q_j}{(z - z_j)}\right)  \right]
\left[\left(\frac{2\pi \bar{z}}{\beta} \right)^2\sum_{j=1}^k\left(\frac{h_j}{(\bar{z} - \bar{z}_j)^2}+ \frac{1}{\bar{z}-\bar{z}_j}\partial_{\bar{z}_j} \right) - \frac{\pi^2 n m c}{6\beta^2} \right]\langle \dots \rangle_{\mathbb{C}}.
\end{dmath}
The twist fields are not charged under the $U(1)$ (see Appendix \ref{U1_charge_twist}), so the leading term is trivial. The same can be said for the $T\bar{J}$ deformation. We found this to also be the case holographically in Section \ref{sec_lin_dilaton}.

\section{Discussion}
\label{sec_discussion}

In this work, we have studied the mutual information and reflected entropy in holographic $T\bar T$ deformed two dimensional quantum field theories. We also studied these entanglement measures perturbatively in $T{\bar J}$ and $J \bar T$ deformations. The results of our calculations have led to several notable physical phenomena. There are also several interesting directions to take from here. 

For $\mu > 0$ and asymptotically linear dilaton geometries, we have found the mutual information and reflected entropy of disjoint intervals to diverge when the intervals are a finite, regulator independent distance ($l_{\min}$) away from each other. Such a phenomenon never occurs in local quantum field theory and signals a breakdown of locality, specifically a breakdown of the split property. In particular, the mutual information \cite{2015JHEP...10..003C} and reflected entropy \cite{2019arXiv190500577D} have been proposed as valuable regularization scheme independent ``geometric regulators'' that can be studied to extract $c$-functions in general dimensions. These schemes rely on taking the ratio between the characteristic size of the regions and the spatial distance between regions to zero. Clearly, this breaks down when we hit the non-locality scale $l_{\min}$. It would be fascinating to consider more carefully how to incorporate non-local theories, such as the ones we have studied, in renormalization group analysis.

For $\mu < 0$ and cutoff AdS geometries, we have found the opposite effect of the theory becoming non-local. Rather than enhanced divergences, we have found all divergences in the mutual information and reflected entropy to be tamed even when the distance between intervals goes to zero. The square root of the deformation parameter acts akin to a UV cutoff. This phenomenon is also never seen in local quantum field theory and is more reminiscent of finite dimensional lattice models. We were able to compare the bulk and boundary computations of the reflected entropy of disjoint intervals at finite temperature perturbatively. While we found the corrections to be formally equivalent, the numerical coefficients were distinct. There are two possible explanations for this tension. Either, non-perturbative corrections save the day as in Ref.~\cite{2020JHEP...04..152L}\footnote{In this analysis, taking the correct order of limits also played a central role.} or modifications must be made to have a consistent duality between the deformed field theory and cutoff AdS. Such modifications may be analogous to the ones shown to be necessary for the inclusion of matter fields in Ref.~\cite{Kraus:2018xrn,2019arXiv190611251G}. We believe resolving this tension is an important direction for future work.

It is also interesting to consider a recent holographic proposal that we have largely neglected in this work where the dual of the $T\bar T$ deformation is proposed to be an ensemble of AdS spacetimes with randomly fluctuating boundary diffeomorphisms \cite{2020arXiv200306300H}. Such a proposal is a natural combination of Cardy's interpretation of the $T\bar T$ deformation as random geometry \cite{2018JHEP...10..186C} and the standard GKPW dictionary of AdS/CFT \cite{1998AdTMP...2..253W,1998PhLB..428..105G}. We see no reason why bulk and boundary computations of mutual information and reflected entropy in this proposal should not match \textit{exactly}, though the computations may be quite technical.  
Confirming this would certainly be worthwhile.

Lastly, we comment on a somewhat tangential future direction. The twist fields needed for the computation for R\'enyi entropies and reflected entropy are uncharged under $U(1)$, leading to somewhat mundane results in perturbation theory for the $J\bar{T}$ and $T\bar{J}$ deformations. However, there are finer grained probes of charged states, such as the symmetry-resolved entanglement \cite{2018PhRvL.120t0602G}, whose corresponding twist fields are charged under $U(1)$. The symmetry-resolved entanglement explains the contribution to the von Neumann entropy from each charge sector. These will presumably have corrections at leading order in $\mu_{\pm}$ and it would be fascinating to understand how these contributions are effected under the charged deformations. 

\acknowledgments
We thank Tom Faulkner, Christian Ferko, David Kutasov, Shinsei Ryu, Gautam Satishchandran, Sav Sethi, Ronak Soni, and Tadashi Takayanagi for useful discussions and comments. MA is supported in part by DOE grant de-sc0009924. JKF is supported through a Simons Investigator Award to Shinsei Ryu from the Simons Foundation.

\appendix

\section{Derivation of holographic entanglement entropy in string frame}
\label{app_string_frame}

Under reasonable assumptions, the Ryu-Takayanagi formula has been derived for theories with arbitrary matter content in Einstein frame\footnote{This is because in Einstein frame, the contribution to the on-shell bulk action for all matter fields is proportional to $n$ (the replica number) when performing the gravitational replica trick while the metric term is proportional to $(n-1)$. We thank Tadashi Takayanagi for explaining this to us.}
\begin{equation}
    S = \frac{1}{4G_N^{(d)}}\int_{\gamma_A} d^{d-2}x \sqrt{g^{(E)}}.
    \label{RT_einstein}
\end{equation}
We would like to know what the analog is in string frame. The relation between the string and Einstein frame metrics is
\begin{equation}
    g_{\mu \nu}^{(s)} = e^{\Phi/2}g_{\mu \nu}^{(E)}.
\end{equation}
Thus, the determinants of the metrics are related as
\begin{equation}
    g^{(s)} = e^{(d-2) \Phi/2}g^{(E)}.
\end{equation}
The reason why we have $(d-2)$ is because we are working with the induced metrics on  codimension-two surfaces. Thus, the RT formula becomes
\begin{equation}
    S = \frac{1}{4G_N^{(d)}}\int_{\gamma_A} d^{d-2}x  e^{-(d-2) \Phi/4}\sqrt{g^{(s)}}.
\end{equation}
Now, let us specify to the $d=10$ supergravity theory that we are generally concerned with
\begin{equation}
    S = \frac{1}{4G_N^{(10)}}\int_{\gamma_A} d^{8}x  e^{-2 \Phi}\sqrt{g^{(s)}}.
    \label{RT_string}
\end{equation} 
This is the formula posited in Ref.~\cite{2006JHEP...08..045R}. It is important to note that the extremal surface in (\ref{RT_string}) is not described by the same coordinates as the one in (\ref{RT_einstein}), though it is the same surface. In the string frame coordinates, it is extremal with respect to the integrand including the dilaton prefactor.

\section{U(1) charges of twist operators}
\label{U1_charge_twist}

In order to evaluate the conformal blocks, we must find the charges under the $U(1)$ symmetry for the twist operators. Let us warm up by computing these for the twist operators used for entanglement entropy, $\sigma_n$, labeled by a single replica index. These have conformal dimensions
\begin{align}
    h_n = \bar{h}_n = \frac{c}{24}\left(n- \frac{1}{n} \right).
\end{align}
These can be determined by considering the $n$-sheeted branched cover of the complex plane, $\mathcal{R}_n$, used to compute the $n^{th}$ power of the reduced density matrix of a single interval, $(u, v)$. We know that on the complex plane, the one-point function of the $U(1)$ current is trivial by rotational and translational invariance
\begin{align}
    \langle J(z) \rangle_{\mathbb{C}} = 0.
\end{align}
$J(z)$ is a chiral primary field of dimension $(1,0)$, so it transforms covariantly under conformal maps
\begin{align}
    J(w) = \frac{\partial z}{\partial w} J(z).
\end{align}
In particular, we consider the conformal map that takes $\mathcal{R}_n$ to $\mathbb{C}$
\begin{align}
    z = \left(\frac{w-u}{w-v}\right)^{1/n}.
\end{align}
This means that $\langle J(w) \rangle_{\mathcal{R}_n} = 0$. We now compare this with the Ward identity for Kac-Moody symmetries
\begin{align}
    \langle J(z) \sigma_n(w_1) \bar{\sigma}_n(w_2)\rangle = \left(\frac{q_1}{z-w_1} + \frac{q_2}{z-w_2 } \right)\langle \sigma_n(w_1) \bar{\sigma}_n(w_2)\rangle.
\end{align}
For this to equal zero for any value of $w_1$ and $w_2$, we must have $q_1 = q_2 = 0$, so the twist fields are neutral under the $U(1)$. 

Now let's proceed to the generalized twist operators \eqref{SR_twist_dim}. The simplest to study is the two point function $\langle \sigma_{g^{\ }_B g_A^{-1}} \sigma_{g^{\ }_A g_B^{-1}}  \rangle $ because the manifold this describes decouples into two copies of $\mathcal{R}_n$, so running through the same argument, we find these operators are also uncharged. This immediately tells us that the two operators that fuse to form them must have opposite charge. We can fix these by considering $\langle \sigma_{g^{\ }_A } \sigma_{g_A^{-1}}  \rangle $ which is needed for computing the reflected entropy between $A$ and the empty set. This factorizes into $n$ $\mathcal{R}_m$'s, so the same argument goes through and we conclude that all twist operators are uncharged under the $U(1)$. This means that the $U(1)$ contribution to the conformal block, \eqref{U1_cont_block}, is unity.

\section{Details of perturbative CFT computations}
\label{app_details}

In this appendix, we present a few details of the calculation leading to \eqref{maintext_pert_EW}. The conformal Ward identity implies
\begin{dmath}
    \int_{\mathcal{M}^{(n,m)}} \langle T \bar{T}\rangle_{\mathcal{M}^{(n,m)}} = \frac{1}{\langle \dots\rangle}_{\mathbb{C}} \int_{\mathcal{M} }\frac{1}{nm} \left[\left(\frac{2\pi {z}}{\beta} \right)^2\sum_{j=1}^k\left(\frac{h_j}{({z} - {z}_j)^2}+ \frac{1}{{z}-{z}_j}\partial_{{z}_j} \right) - \frac{\pi^2 n m c}{6\beta^2} \right]
\left[\left(\frac{2\pi \bar{z}}{\beta} \right)^2\sum_{j=1}^k\left(\frac{h_j}{(\bar{z} - \bar{z}_j)^2}+ \frac{1}{\bar{z}-\bar{z}_j}\partial_{\bar{z}_j} \right) - \frac{\pi^2 n m c}{6\beta^2} \right]\langle \dots \rangle_{\mathbb{C}}.
\end{dmath}
The correlation function of four twist operators may be computed using \eqref{HHLL}. 
The total integral becomes
\begin{align}
    &\int_{-\infty}^{\infty} dx \int_{0}^{\beta} d\tau \frac{-\pi ^4 c^2 \mu  (z_1-z_2) (z_3-z_4)}{9 \beta ^4} 
    \nonumber
    \\
    &\times\Bigg[\frac{e^{\frac{4 \pi  (x+i \tau )}{\beta }}}{\left(-z_1+e^{\frac{2 \pi  (x+i \tau )}{\beta
   }}\right) \left(-z_2+e^{\frac{2 \pi  (x+i \tau )}{\beta }}\right) \left(-z_3+e^{\frac{2 \pi  (x+i \tau )}{\beta }}\right) \left(-z_4+e^{\frac{2 \pi  (x+i
   \tau )}{\beta }}\right) \sqrt{\frac{(z_1-z_2) (z_3-z_4)}{(z_1-z_3) (z_2-z_4)}}}
   \nonumber
   \\
   &+\frac{e^{\frac{4 \pi  (x-i \tau )}{\beta
   }}}{\left(-z_1+e^{\frac{2 \pi  (x-i \tau )}{\beta }}\right) \left(-z_2+e^{\frac{2 \pi  (x-i \tau )}{\beta }}\right) \left(-z_3+e^{\frac{2 \pi  (x-i \tau
   )}{\beta }}\right) \left(-z_4+e^{\frac{2 \pi  (x-i \tau )}{\beta }}\right) \sqrt{-\frac{(z_2-z_1) (z_3-z_4)}{(z_1-z_3)
   (z_2-z_4)}}}\Bigg]
\end{align}
We first do the $\tau$ integral to find
\begin{align}
    &\int_{-\infty}^{\infty} dx\frac{i \pi ^3 c^2 \mu  (z_1-z_2) (z_3-z_4)}{18 \beta ^3} \Bigg[-\frac{z_1 \log \left(-z_1+e^{\frac{2 \pi  (x+i \tau )}{\beta
   }}\right)}{(z_1-z_2) (z_1-z_3) (z_1-z_4) \sqrt{\frac{(z_1-z_2) (z_3-z_4)}{(z_1-z_3)
   (z_2-z_4)}}}
   \nonumber
   \\
   &+
   \frac{z_2 \log \left(-z_2+e^{\frac{2 \pi  (x+i \tau )}{\beta }}\right)}{(z_1-z_2) (z_2-z_3) (z_2-z_4)
   \sqrt{\frac{(z_1-z_2) (z_3-z_4)}{(z_1-z_3) (z_2-z_4)}}}
   +
   \frac{z_3 \log \left(-z_3+e^{\frac{2 \pi  (x+i \tau
   )}{\beta }}\right)}{(z_1-z_3) (z_3-z_2) (z_3-z_4) \sqrt{\frac{(z_1-z_2) (z_3-z_4)}{(z_1-z_3)
   (z_2-z_4)}}}
   \nonumber
   \\
   &-
   \frac{z_4 \log \left(-z_4+e^{\frac{2 \pi  (x+i \tau )}{\beta }}\right)}{(z_1-z_4) (z_2-z_4) (z_4-z_3)
   \sqrt{\frac{(z_1-z_2) (z_3-z_4)}{(z_1-z_3) (z_2-z_4)}}}
   +
   \frac{z_1 \log \left(-z_1+e^{\frac{2 \pi  (x-i \tau
   )}{\beta }}\right)}{(z_1-z_2) (z_1-z_3) (z_1-z_4) \sqrt{-\frac{(z_2-z_1) (z_3-z_4)}{(z_1-z_3)
   (z_2-z_4)}}}
   \nonumber
   \\
   &+
   \frac{z_2 \log \left(-z_2+e^{\frac{2 \pi  (x-i \tau )}{\beta }}\right)}{(z_1-z_2) (z_3-z_2) (z_2-z_4)
   \sqrt{-\frac{(z_2-z_1) (z_3-z_4)}{(z_1-z_3) (z_2-z_4)}}}
   -
   \frac{z_3 \log \left(-z_3+e^{\frac{2 \pi  (x-i \tau
   )}{\beta }}\right)}{(z_1-z_3) (z_3-z_2) (z_3-z_4) \sqrt{-\frac{(z_2-z_1) (z_3-z_4)}{(z_1-z_3)
   (z_2-z_4)}}}
   \nonumber
   \\
   &+
   \frac{z_4 \log \left(-z_4+e^{\frac{2 \pi  (x-i \tau )}{\beta }}\right)}{(z_4-z_1) (z_4-z_2) (z_4-z_3)
   \sqrt{-\frac{(z_2-z_1) (z_3-z_4)}{(z_1-z_3) (z_2-z_4)}}}\Bigg]\Bigg|_{\tau = 0}^{\tau = \beta}
\end{align}

Every term that either does not depend on $\tau$ or depends on $\tau$ only exponentially and not within a logarithm is trivial when evaluating the difference of the indefinite integral at $\beta$ and $0$. The terms with the exponential within the logarithm must be treated with care. This has been discussed in e.g.~Ref.~\cite{2018PhRvD..98h6025C}. Due to the branch cut, we have
\begin{align}
    \log \left(e^{\frac{2\pi (x+ i \tau)}{\beta}} - e^{\frac{2\pi l}{\beta}} \right)\Big|_{\tau = 0}^{\tau = \beta} = \begin{cases}
    0, & x < l
    \\
    2\pi i, & x > l
    \end{cases}.
\end{align}
Analogously, for the complex conjugate, we run around the branch cut the opposite direction, so
\begin{align}
    \log \left(e^{\frac{2\pi (x- i \tau)}{\beta}} - e^{\frac{2\pi l}{\beta}} \right)\Big|_{\tau = 0}^{\tau = \beta} = \begin{cases}
    0, & x < l
    \\
    -2\pi i, & x > l
    \end{cases}.
\end{align}
After evaluating, we find \eqref{maintext_pert_EW}.

\section{Non-perturbative CFT calculation for single-trace \texorpdfstring{$T\bar{T}$}{TEXT}}
\label{app_nonpert}

In this section, we repeat the analysis done in Ref.~\cite{2018PhRvL.121m1602D} for the entanglement entropy of a region with an entangling surface of antipodal points on $S^2$ except for the single-trace deformation dual to the asymptotically linear dilaton geometry. Here, we assume that the spacetime conformal field theory is a symmetric product orbifold $\mathcal{M}^N/S_N$. Each block has central charge $\tilde{c}$, so the total CFT has central charge $c = N\tilde{c}$. We apply the replica trick by considering the $n$-sheeted cover of the sphere of radius $r$
\begin{align}
    ds^2 = r^2 \left( d\theta^2+n^2\sin \theta d\phi^2\right).
\end{align}
The von Neumann entropy is then
\begin{align}
    S = \left(1 - n \frac{\partial }{\partial n} \right) \log \mathcal{Z}|_{n = 1}.
\end{align}
The sphere partition function responds to a change in $n$ as
\begin{align}
    \frac{\partial \log \mathcal{Z}}{\partial n}\Big|_{n = 1} = -\frac{1}{2}\int\sqrt{g}T,
\end{align}
where $T$ is the trace of the stress tensor. Similarly, when $n=1$, the response of the sphere partition function to a change in radius is
\begin{align}
    \frac{d \log \mathcal{Z}}{dr} = -\frac{1}{r}\int \sqrt{g} T,
\end{align}
so the entropy may be rewritten as
\begin{align}
    S = \left( 1- \frac{r}{2}\frac{\partial}{\partial r}\right) \log \mathcal{Z} .
\end{align}
The trace of the stress tensor flows in a known way under a $T\bar{T}$ deformation
\begin{align}
    \langle T^a_a \rangle = -\frac{\tilde{c}}{24\pi} R-\frac{\tilde{\mu}}{4}\left(\langle T^{ab}\rangle \langle T_{ab}\rangle - \langle T^a_a\rangle^2 \right).
\end{align}
However, in the symmetric orbifold theory, we have
\begin{align}
    T_{ab} = \sum_i^N T_{a b}^{(i)},
\end{align}
with each individual copy flowing under a $T\bar{T}$ deformation. Thus, there is an extra factor of $N $
\begin{align}
    \langle T^a_a \rangle = -\frac{c}{24\pi} R-\frac{ \mu}{4}\left(\langle T^{ab}\rangle \langle T_{ab}\rangle - \langle T^a_a\rangle^2 \right).
    \label{flow_eq}
\end{align}
We have absorbed the factors of $N$ into $c$ and $\mu$ due to our definitions of the central charge and deformation parameter. By symmetry and \eqref{flow_eq}, the stress tensor must be proportional to the metric as
\begin{align}
    T_{ab} = \frac{2}{ \mu}\left(1- \sqrt{1 +\frac{c \mu }{24\pi r^2}} \right)g_{ab},
\end{align}
so we have
\begin{align}
    \frac{\partial \log \mathcal{Z}}{\partial r} = \frac{16\pi }{ \mu }\left( \sqrt{r^2+\frac{c\mu}{24\pi}}-r\right).
\end{align}
Imposing the boundary condition $\log \mathcal{Z}(r=0) = 0$, we have the entropy
\begin{align}
    S = \frac{c}{3}\sinh^{-1}\left(\sqrt{\frac{24\pi}{c \mu}}r \right).
\end{align}
This is identical to the double-trace formula from Ref.~\cite{2018PhRvL.121m1602D}. It would be interesting to check this holographically by finding the Euclidean compactification of \eqref{lin_dil_metric} to a sphere. However, the boundary condition for the differential equation would need to be appropriately modified because, for the relevant sign of $\mu$, we expect the partition function to diverge at a finite value of $r$.  



\providecommand{\href}[2]{#2}\begingroup\raggedright\endgroup

\end{document}